\documentclass[twocolumn,traditabstract]{aa}
\usepackage{natbib}
\newcommand{\bibfont}{\tiny}
\usepackage{graphicx}
\usepackage[T1]{fontenc}
\usepackage{amsmath}
\usepackage{amssymb}

\usepackage{amsmath}
\usepackage{graphicx}
\usepackage{amsmath}
\usepackage{amssymb}
\usepackage{amsfonts}
\usepackage{graphicx,epsfig,graphics}
\usepackage{aas_macros}
\usepackage{graphics}
 
\def\be{\begin{eqnarray}}
\def\ee{\end{eqnarray}}

\graphicspath{{}}

\title{Sparse component separation for accurate CMB map estimation}
\author{J. Bobin, J.-L. Starck, F. Sureau and S. Basak}
\institute{Laboratoire AIM, UMR CEA-CNRS-Paris 7, Irfu, SAp/SEDI, Service d'Astrophysique, CEA Saclay, F-91191 GIF-SUR-YVETTE CEDEX, France.}

\begin{document}
 

\abstract{
The Cosmological Microwave Background (CMB) is of premier importance for the cosmologists to study the birth of our universe. Unfortunately, most CMB experiments such as COBE, WMAP or Planck do not provide a direct measure of the cosmological signal; CMB is mixed up with galactic foregrounds and point sources. For the sake of scientific exploitation, measuring the CMB requires extracting several different astrophysical components (CMB, Sunyaev-Zel'dovich clusters, galactic dust) form multi-wavelength observations. Mathematically speaking, the problem of disentangling the CMB map from the galactic foregrounds amounts to a component or source separation problem. In the field of CMB studies, a very large range of source separation methods have been applied which all differ from each other in the way they model the data and the criteria they rely on to separate components. Two main difficulties are i) the instrument's beam varies across frequencies and ii) the emission laws of most astrophysical components vary across pixels.
This paper aims at introducing a very accurate modeling of CMB data, based on sparsity,  accounting for beams variability across frequencies as well as spatial variations of the components' spectral characteristics. 
Based on this new sparse modeling of the data, a sparsity-based component separation method coined Local-Generalized Morphological Component Analysis (L-GMCA) is described. Extensive numerical experiments have been carried out with simulated Planck data. These experiments show the high efficiency of the proposed component separation methods to estimate a clean CMB map with a very low foreground contamination, which makes L-GMCA of prime interest for CMB studies.}


\date{Received -; accepted -}

\maketitle

\section*{Introduction}
Cosmological Microwave Background is generally not observable directly; most CMB experiments such as WMAP or Planck provide multi-wavelength observations in which the CMB is mixed with other astrophysical components. Recovering useful cosmological information requires disentangling in the CMB data the contribution of several astrophysical components namely CMB itself, Galactic emissions from thermal dust, 
spinning dust and synchrotron, Sunyaev-Zel'dovich (SZ) clusters, point sources, free-free emission, CO emission to only name a few, see \citet{Bouchet}.\\
The recovery of these sources falls in the framework of component separation. The CMB itself dominates over a large fraction of the sky at frequencies lower than $100$ GHz, but for an efficient component separation a range of channels needs to be observed. Component separation consists of estimating a set of parameters which describe the components of interest as well as how they contribute to the data. For example, it could be parameters describing the statistical properties such as power spectra, electromagnetic spectra to only name two.\\
A large range of component separation techniques have been proposed; they mainly differ from each other in the way they model the data and the assumptions made on the components to be separated. However, none of state-of-the-art component separation techniques accurately models CMB data. Such a modeling is challenging as it involves a wide variety of instrumental and physical phenomena such as~:
\begin{itemize}
\item{\bf Instrumental noise~:} It is generally correlated and non stationary~: each position in the sky is not observed the same number of times. The noise variance varies across pixels.\\
\item{\bf Point sources~:} point sources are very hard to account for in component separation~: each point source has its own electromagnetic spectrum. Therefore the contribution of point sources can't be defined as a simple spatial template that scales across frequencies.\\ 
\item{\bf Emissivity variations~:} It is a well established fact that the emissivity of most foregrounds such as dust or synchrotron varies across the sky. Again, assuming that the electromagnetic spectrum of these components is constant across pixels is inaccurate.\\
\item{\bf Heterogeneous beams~:} the observed maps have generally different resolutions; furthermore, the beams are not necessarily isotropic and spatially invariant.\\
\end{itemize}
Each observed channel $x_i$, at frequency $\nu_i$, contains information about several sky emissions (CMB, dust, etc) which can be written at each pixel $k$ as follows~:
\begin{eqnarray}
x_i [k] = b_i \star \left( \sum_{j=1}^{N_s}  a_{i, j}[k] s_j[k] \right)+ n_{i}[k]
\label{eq_mixte}
\end{eqnarray}
where $i$ is the channel number,  $k$ is the pixel position, $N_s$ the number of sky emissions, $b_{i}$ is the instrument beam ({\it i.e.} point spread function), $a_{i, j}[k]$ models for the contribution of source $j$ in channel $i$, $s_{j}$ is  the $j$-th sky component and $n_{i}[k]$ is the noise.\\
A first - commonly found - approach to get the CMB, called template fitting, consists in fitting sky templates $t_j$ (for all non-CMB sky emissions).
But many other strategies have been investigated in the last years in order to extract the best CMB map from WMAP or Planck data.
If we omit the effect of the beams and assume that the sky component spectral behavior does not vary across the sky (i.e. $ a_{i,j}[k]$  is  constant for any fixed $(i,j)$), then the above equation can be recast as~:
\begin{eqnarray}
\label{eq:GMCA1}
{\bf X} = {\bf A S} + {\bf N}
\end{eqnarray}
Where ${\bf X}$ are the observed multichannel data, ${\bf S}$ the unknown sources, and ${\bf A}$ is the unknown or partially unknown mixing matrix
($a_{i,j}$ is the contribution of component $s_{j}$ to the channel $i$ at frequency $\nu_i$). Therefore, one needs to find ${\bf S}$ and ${\bf A}$ from the observed data ${\bf X}$ only.
This is obviously a highly ill-posed problem known, in the statistics community, as Blind Source Separation (BSS).
During the last decades various methods have been introduced to tackle BSS problems. The main differences between all these component separation methods are the statistical assumptions made to differentiate between the sources. A famous example is Independent Component Analysis which rely on the statistical independence of the sources.  Although independence is a strong assumption, it is in many cases physically acceptable, and provides much better solutions than using a simple second order decorelation assumption generally obtained with methods such as using Principal Component Analysis (PCA).\\
Another example is Internal Linear Combination (ILC), which is not \textit{stricto sensu} a component separation but rather a component extraction method in the sense that it only estimates a single component with assumed known spectrum. In statistics, ILC is closely linked to the "Best Linear Unbiased Estimator" (\textit{a.k.a.} BLUE) with the assumption that the covariance of the error is identical to the covariance of the data. The most basic ILC method amounts to applying this estimation formula to the data in the pixel domain. Further refinements have been introduced in \citet{NeedletILC} where the ILC estimator of the CMB is computed in local patches in the wavelet space using so-called Needlet filters\footnote{Like undecimated wavelets on the sphere \citep{starck:sta05_2}, the needlet signal representation is a tight frame computed via filter banks in spherical harmonics.}.\\
Beyond statistical independence, the Generalized Morphological Component Analysis method (GMCA) \citep{bobin:gmca_itip,bobin-gmca-cmb} makes profit of the fact that most foreground emissions can be sparsely represented in a well chosen signal representation (\textit{e.g.} wavelets). The estimation of the components and the unknown coefficients in the mixing matrix is performed by enforcing the sparsity of the components in the wavelet domain.\\
In addition, parametric methods such as MEM \citet{FastMEM_98} and Commander \citet{Commander} have been considered, based on a full physical modeling of the sky components.\\

Except for GMCA\footnote{GMCA is part of the ISAP toolbox~: {\it http://jstarck.free.fr/isap.html}.}, codes relative to these different methods are not public. This makes the comparison between them relatively difficult. However, a first comparison of these methods has been done in \citet{Leach_08} on full-sky simulated Planck data. Whatever the kind of approach one adopts, efficient CMB estimation requires an accurate modeling of the data as well as a robust and effective separation technique. Also, with the future Planck release in $2013$, we need to have a much better understanding of what methods work better to recover a high quality CMB map from full-sky surveys.\\ \\

\textit{Content of this paper~:}
In this paper, we first review in Section~\ref{review_cmb_compsep} the major classes of component separation methods, and we discuss the advantages and drawbacks of each of them. Section~\ref{subsec:sparsity} discusses the use of sparsity prior for CMB estimation. Then we show in section~\ref{sec:lgmca} how the GMCA method can be modified in order to properly take into account the different resolutions of the different channels, and the spatial variation of the mixing matrix.
Based on this sophisticated data modeling, a novel sparse component separation method Local GMCA (L-GMCA) coined is described. Finally, results of extensive numerical experiments are presented in Section~\ref{sec:results}, which show
the advantage on the sparse recovery method over the ILC-based approach.

\section{CMB Recovery}
\label{review_cmb_compsep}

Components separation techniques can be grouped in a few classes according to the criterion used for separating the sources. In this section we present the various methods proposed for CMB recovery and discuss their relative advantages and drawbacks.

\subsection{Template Fitting}

Having the foregrounds as additional components of the microwave sky,
one can perform a fit of the template to the data for foreground analysis.
For $N_t$ template vectors, $\{t_i\}_{i=1,\cdots,N_T}$, the template-corrected data has the form at frequency $\nu_i$
\begin{equation}
\tilde{x}_i = x_i-\sum_{j = 1}^{N_t} \beta_{j} t_{j}\;,
\end{equation}
where the best-fit amplitude, $\beta_{j}$, for each foreground template
can be obtained by minimizing $\tilde{x}_i^{T}\boldsymbol{C}^{-1}\tilde{x}_i$,
where $\boldsymbol{C}$ is the total covariance matrix for the template-corrected
data $\boldsymbol{C}=\left\langle \tilde{x}_i\tilde{x}_i^{T}\right\rangle $.
Template-fitting can be performed either in the pixel domain or in the harmonic space. Pixel-based implementation allows for incomplete sky coverage as well as a refined modeling of non-stationary noise at the expense of a more complex modeling of the data~: the CMB covariance matrix is large and dense. Conversely, spherical harmonics make the modeling of CMB much simpler but requires crude stationary assumptions for noise.\\

CMB cleaning by template fitting has a number of advantages, the first to be its
simplicity. The technique makes full use of the spatial information
in the template map, which is important for the non-stationary, highly
non-Gaussian emission distribution, typical of Galactic foregrounds. xThere are also
disadvantages to this technique~: the noise of the template is added
to the solution, and imperfect models
of the templates could add systematics and non-Gaussianities to the
data. Template fitting also assumes that the electromagnetic spectrum of the emission
related to a given template is not varying spatially, which is generally not true.

Refer to \citet{Dunkley_Template} for a more detailed description of template
fitting techniques.\\
Template cleaning of the COBE/FIRAS data reduced a complicated foreground
by a factor of $10$ by using only $3$ spatial templates \citep{Fixsen_TemplateCOBE}. 

\subsection{Second order statistical methods}

\subsubsection*{ILC: Internal Linear Combination}

In the framework of ILC, very little is assumed about the
different components to be separated out. The main component is assumed
to be the same in all the frequency bands and the observations
are calibrated with respect to this component. Each observation $x_i$ is modeled at pixel $k$ as follows~:
\begin{equation}
x_{i}[k]=s[k]+f_{i}[k]+n_{i}[k]\:,
\end{equation}
where $i$ denotes the frequency channels at frequency $\nu_i$, $f_{i}[k]$ and $n_{i}[k]$
are the foregrounds and noise contributions at pixel $k$ respectively.
One then looks for the solution~:
\begin{equation}
\hat{s}[k]=\sum_{i}w_{i}[k]x_{i}[k]\:,
\end{equation}
The simplest version of ILC assumes the weights $w_{i}[k]$ are constant across the sky and therefore are not function of $k$. The ILC estimate of the CMB is then obtained by estimating the weights $w_i$ which minimize the least square of the residual $x_i  - s$ for all $i$. Assuming that the covariance matrix of the error the is covariance matrix of the data themselves (this assumption is a good approximation when the foregrounds and/or intrumental noise are preeminent)~:
\begin{equation}
\label{eq:ILC}
\hat{s} = \min_s \left \| {\bf X} - {a}^T s \right \|_{F,{\bf \Sigma_{X}}}^2
\end{equation}
where the norm $\| \, . \, \|_{F,{\bf \Sigma_{X}}}$ stands for the weighted Frobenius norm~:$\| {\bf Y} \|_{F,{\bf \Sigma_{X}}} = \mbox{Trace}\left({\bf Y}^T{\bf \Sigma_{X}}^{-1} {\bf Y}\right)$ and $a$ is the electromagnetic spectrum of CMB (it is a vector of ones for data in thermodynamic units). Solving the minimization problem in Equation~\ref{eq:ILC} leads to~: 
$$
\hat{s} = \frac{a^T {\bf \Sigma_{X}}^{-1}}{a^T {\bf \Sigma_{X}}^{-1}a} {\bf X}
$$
Note that the ILC yields the minimization the \textit{{total}}
variance of the ILC map. Away from the galactic plane and on small scales,
the best linear combination for cleaning the CMB map from foregrounds
and noise might be different from regions close to the galactic plane
or the large scales. To improve on this, the map is generally decomposed into several
regions and ILC is applied to them independently. The ILC performs well when no prior information is known about the different components~: the only prior knowledge is the CMB electromagnetic spectrum. Therefore,
ILC is considered as a foreground removal technique rather than a component separation technique.

\subsubsection*{Correlated Component Analysis (CCA)}

This method \citep{CCA_Bedini} is a semi-blind approach that estimates
the mixing matrix on sub-patches of the sky based on second order
statistics. It makes no assumptions about the independence
of the sources. This method adopts the commonly used models for the
sources to reduce the number of parameters estimated and exploits the
spatial structure of the source maps. The spatial structure of the
maps are accounted for through the covariance matrices at different
shifts $(\tau,\psi$)\begin{eqnarray}
\boldsymbol{C}_{d}(\tau,\psi) & = & \boldsymbol{A}\boldsymbol{C}_{s}(\tau,\psi)\boldsymbol{A}^{t}+\boldsymbol{C}_{n}(\tau,\psi)\;,\end{eqnarray}
where $\boldsymbol{C}_{d}(\tau,\psi)$ is estimated from data and
the noise covariance matrix $\boldsymbol{C}_{n}(\tau,\psi)$ is derived
from the map-making noise estimations. Then by minimizing the equation
\begin{equation}
\sum_{\tau,\psi}\left\Vert \boldsymbol{A}\boldsymbol{C}_{s}(\tau,\psi)\boldsymbol{A}^{t}-\left[\boldsymbol{C}_{d}(\tau,\psi)-\boldsymbol{C}_{n}(\tau,\psi)\right]\right\Vert \;,\end{equation}
where the Frobenius norm is used, one can estimate the mixing matrix
and the free parameters of the source covariance matrix. Given as
estimate of $\boldsymbol{C}_{s}$ and $\boldsymbol{C}_{n}$, the above
equation can be inverted and component maps obtained via the standard
inversion techniques of Wiener filtering or generalized least square
inversion. To obtain a continuous distribution of the free parameters
of the mixing matrix, CCA is applied to a large number of partially
overlapping patches. 

\subsubsection*{Spectral Matching ICA (SMICA)}

SMICA \citep{ica:Del2003} is a ICA-based components separation technique relying on second order statistics
in the frequency or spherical harmonics domain. For multichannel maps $x_{i}[k]$ one
computes
\begin{equation}
\hat{\boldsymbol{R}}_{\ell}=\frac{1}{2\ell+1}\sum_{m}\boldsymbol{x}_{\ell m}\boldsymbol{x}_{\ell m}^{\mathrm{\dagger}}\:,
\end{equation}
for each $\ell$ and $m$. One then models the ensemble-average as
$\boldsymbol{R}_{\ell}=\left\langle \hat{\boldsymbol{R}}_{\ell}\right\rangle =\sum_{j}\boldsymbol{R}_{\ell}^{j}$
where the sum is over the components. For each component, $\boldsymbol{R}{}_{\ell}^{j}$
is a function of a parameter vector $\theta^{j}$, where the parameterization
embodies the prior knowledge about the components as well as the mixing matrix. The parameters
are determined by minimizing the \textit{spectral mismatch} \begin{equation}
\textrm{min}_{\theta}\sum_{\ell}(2\ell+1)K(\hat{\boldsymbol{R}}_{\ell}|\sum_{j}\boldsymbol{R}_{\ell}^{j})\:,\end{equation}
where $K(\boldsymbol{C}_{1}|\boldsymbol{C}_{2})$ is a measure of
mismatch between $\boldsymbol{C}_{1}$ and $\boldsymbol{C}_{2}$ (typically the Kullback-Leibler divergence between two Gaussian distributions with same mean and covariance matrices $\hat{R}_{\ell}$ and $\sum_j R_{\ell}^j$). 

\subsection{Parametric methods}
\subsubsection*{Maximum Entropy Method (MEM)}

Having a hypothesis $H$ in which the observed data
$\{x_i\}_{i=1,\cdots,M}$ are a function of the underlying signal $s$ one can follow
Bayes' theorem, which tells us the posterior probability is the product
of the likelihood and the prior probability divided by the evidence;
\begin{equation}
P(s|d,H)=\frac{P(d|s,H)P(s|H)}{P(d|H)}\:.
\end{equation}
Then following the maximum entropy principle, one
uses a prior distribution which maximizes the entropy of the estimate given a set of
constraints \citet{FastMEM_98}.
 The MEM can be implemented in the spherical harmonic domain
where the separation is performed mode-by-mode which speeds up the
optimization. Based on this maximum entropy approach, FastMEM is a non-blind
method, which means that the spectral behavior of the components must
be known beforehand. Further details of this method are presented in \citet{FastMEM_2005}.

\subsection{Comparisons of assumptions and modeling}

These different methods give a rather large overview of the different approaches used so far to estimate the CMB from multi-wavelength observations. They all differ in the way they model the data and the assumptions made to disentangle the different components~:
\begin{itemize}
\item{\bf Instrumental noise~:} only pixel-based or wavelet-based methods are able to, at least approximately, model for the variation across pixels of the noise variance. Methods based on spherical harmonics rely only on the power spectra or on cross spectra to perform the separation; instrumental noise can be accounted for via its power spectrum which do not give a precise characterization of its non-stationarity.\\
\item{\bf Frequency beams~:} the beam of the observations varies across frequencies. The effect of the beam being a simple multipole-wise product in spherical harmonics, methods performing in this domain can effectively model for its effect. Pixel-based methods generally neglect the variation of the beam across frequency or at least estimate the mixture parameters ({\it i.e.} mixing matrix) at a common low resolution at the expense of the lost of small scale information.\\
\item{\bf Component's modeling~:} Parametric methods such as FastMEM \citet{FastMEM_98} or Commander \citet{Commander} are attempts to make use of accurate modeling of astrophysical components involving spatially variant parameters ({\it e.g.} spectral index and temperature of dust). The dimension of the parameter space growing with the resolution, parameters are generally estimated at low resolution and extrapolated to higher resolution. If this allows for a precise modeling of foregrounds at large scale, this model is still inaccurate to capture small scale variations of the components. BSS-based methods such as CCA \citep{CCA_Bedini} have also been extended so as to incorporate a parametric modeling of the major foregrounds (free-free, synchrotron and dust emissions) but with the assumption that these electromagnetic spectrum of these components is fixed across the sky. To our knowledge, only ILC \citet{NeedletILC} in the needlet domain has been extended to perform on patches to allow for space varying ILC weights.\\
\item{\bf Separation criteria~:} When no physics-based assumptions are made on the components, blind source separation techniques such as CCA \citep{CCA_Bedini} or SMICA \citep{ica:Del2003} rely on statistical separation criteria to differentiate between the components. As described in the previous section, CCA and SMICA both make use of second-order statistics to separation components, either covariance matrices in the pixel domain for CCA and ILC or in spherical harmonics for SMICA. If second-order statistics provide sufficient statistics for Gaussian random fields like the CMB, it is no more optimal for non-stationary and non-Gaussian components such as foregrounds. In this, higher-order statistics should also play a preeminent role to measure discrepancies between the sources.\\
\end{itemize}
As emphasized in the introduction, designing a component separation methods allowing for an accurate modeling of the data ({\it i.e.} space-varying noise variance, heterogenous beams) as well as a precise modeling of the data ({\it i.e.} accounting for the space-variant spectral characteristic of components, effective separation criterion for both non-Gaussian foregrounds and CMB) is a challenging task. Up to know, none of the proposed methods takes into consideration all the aspects of data and component modeling.

\subsection{Towards wavelets and sparsity}

The discussion of the previous section sheds light on the respective advantages of data modeling in the pixel space and spherical harmonics. However, it appears clearly that none of these two different approaches appropriately deals with the separation of non-stationary and/or non-Gaussian signals as well as the correlation between pixels of the components\\
Taking the best of both approaches is generally made by switching to a wavelet-based modeling of the data~: i) the wavelet decomposition of the data leads to a splitting of the spherical harmonics domain which allows for a localization in frequency or scale together with a localization in space. Hence, harmonic methods such as SMICA as well as pixel-based techniques such as ILC have been extended with success in the wavelet domain, 
 using isotropic wavelets on the sphere \citep{starck:sta05_2,marinucci08},  leading to WSMICA \citep{SMICA_WT}, N-ILC \citep{NeedletILC} and GenILC \citep{Remazeilles2011} methods. 
 
Owing to the spatial localization of the wavelet representation, CMB estimation is carried out on different regions of the sky in the different wavelet scales which allows for a more effective cleaning of non-stationary components. 
Similarly, a template fitting technique such as SEVEM is now applied in the wavelet domain \citep{WSEVEM} to capture scale-space variabilities of the templates' emissivity. However, this method makes use of Haar wavelets on Healpix faces which is certainly sub-optimal since this specific wavelet function is irregular and exhibits very poor mathematical properties. 
If the data modeling is more complex than a simple template fitting, this approach does not have the versatility of N-ILC and cannot capture variation of the spectral emission of a given component inside a given scale.

\begin{figure}[htb]
\begin{tabular}{c}
\includegraphics[scale=0.2]{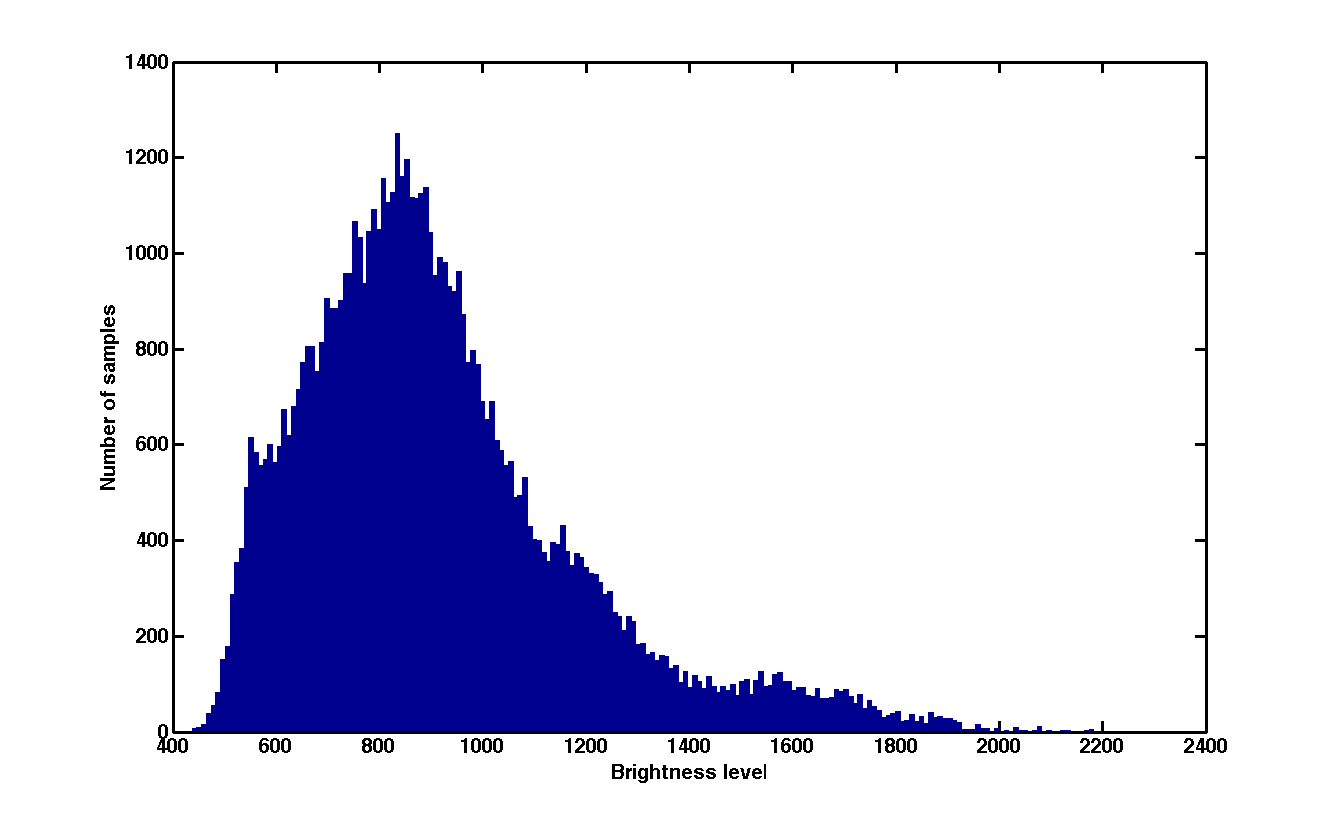} \\
\includegraphics[scale=0.2]{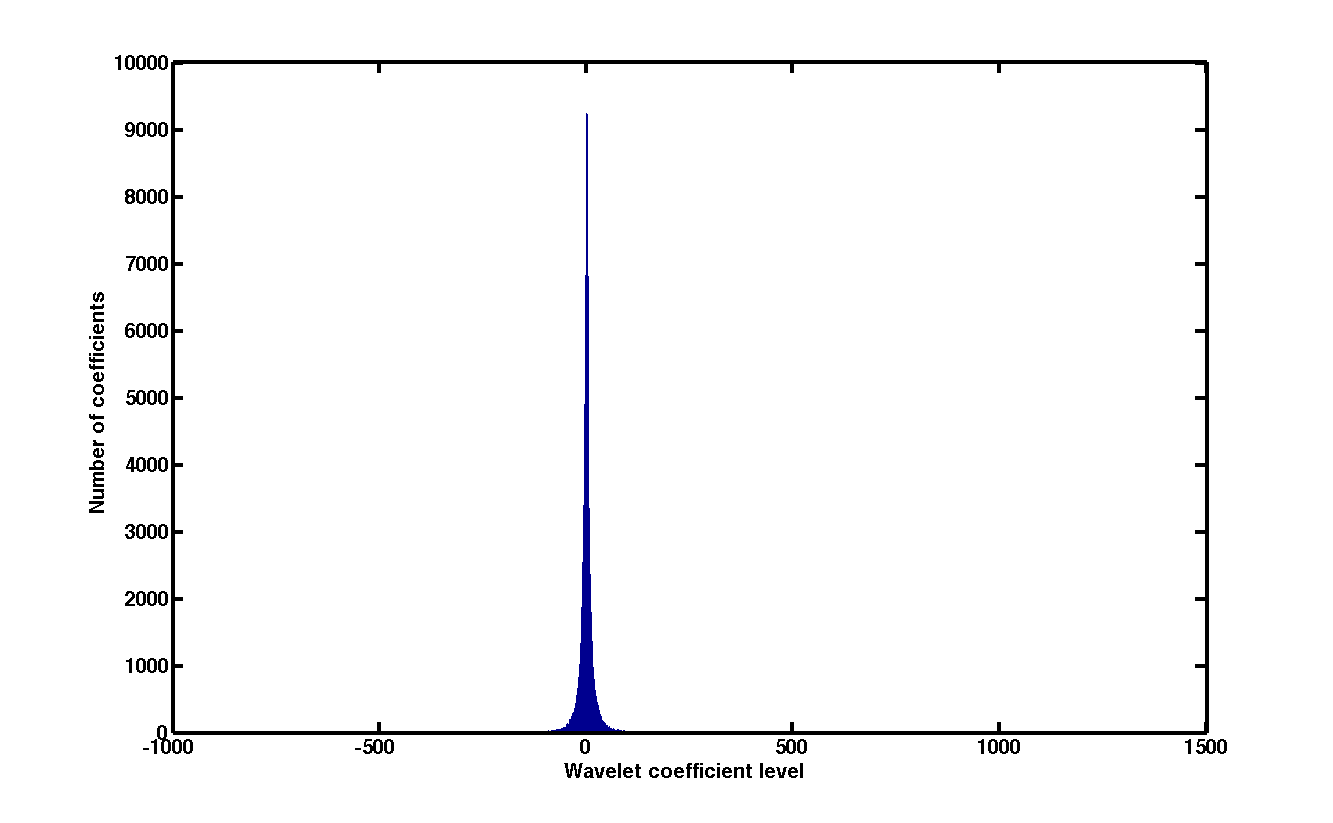} 
\end{tabular}
\caption{Histogram of simulated dust emission at $867$GHz in pixel domain (top), wavelet domain at the (bottom).}
\label{fig:histdust}
\end{figure}

It is important to note that the wavelet transform has the ability to capture coherence or correlation across pixels while averaging the noise contribution. This essential property is also known in the field of modern statistics as sparsity~: correlated structures in the pixel domain are concentrated in a few wavelet coefficients. As an illustration, Figure~\ref{fig:histdust} displays the histogram of the pixel intensity of simulated dust emission at $857$GHz in the pixel domain in the top panel and in the wavelet domain in the bottom panel. This figure particularly shows that if all the pixels of dust are nonzero, the vast majority of its wavelet coefficients are very close to zero and only a few have a significant amplitude. This enlightens the ability of the wavelet transform to concentrate the geometrical content ({\it i.e.} correlation between pixels) of dust emission in a few coefficients.  \\
Extensions to the wavelet domain of the above CMB estimation techniques benefit from the space/frequency localization of the wavelet analysis. However they do not make profit of the sparsity, and thus highly non-Gaussian, property of the wavelet decomposition of the components. Conversely, Generalized Morphological Component Analysis (GMCA) further enforces sparsity to better estimate the sought after sources in the wavelet domain. This component separation method is described in the sequel.

\section{Sparse Component Separation: Generalized Morphological Component Analysis (GMCA)}
\label{subsec:sparsity}

\subsection{Sparsity for component separation}
Sparse priors have been shown to be very useful in regularizing ill-posed inverse problems \citep{starck:book10}. In addition, sparse priors using wavelet bases have been used  with success to various signal processing problems in astronomy including denoising, deconvolution and inpainting \citep{starck:book06}. Like ICA-based techniques, GMCA aims at solving a blind or semi-blind source separation problem. However, GMCA performs in the wavelet domain \citep{bobin-gmca-cmb} to benefit from the sparsity property of the foregrounds in this domain. It is worth mentioning that sparsity has also been used for component separation in fields of research outside 
astrophysics \citep{miki:Zibulevsky,bobin08_aiep,SPIEGMCA09}.\\


In the sequel, we denote by $\boldsymbol{\Phi}$ the forward wavelet transform and $\boldsymbol{\Phi}^T$ its backward transform. Mathematically speaking, this is a matrix made of wavelet waveforms. One can uniquely decompose each source $s_j$ in the wavelet domain as follows~:
$$
s_{j}=\alpha_{j}\boldsymbol{\Phi}
$$
$\alpha_j$ are the expansion coefficients of source $s_j$ in the wavelet basis. The sparsity of the sources means that most of the entries of $\alpha_j$
are equal or very close to zero and only a few has significant amplitudes. The multichannel data $\boldsymbol{X}$ can be written as
\begin{equation}
\boldsymbol{X}=\boldsymbol{A}\mathbf{\alpha}\boldsymbol{\Phi} + \boldsymbol{N}\:,
\label{eq:tensor1-1}
\end{equation}
The objective of GMCA is to seek an unmixing scheme, through the estimation of $\boldsymbol{A}$,
which yields the sparsest sources $\boldsymbol{S}$ in the wavelet domain. This is expressed
by the following optimization problem written in the augmented Lagrangian form~:
\begin{equation}
\label{eq:GMCA2}
\min\frac{1}{2}\left\Vert \boldsymbol{X}-\boldsymbol{A}\mathbf{\alpha}\boldsymbol{\Phi}\right\Vert _{F}^{2}+\lambda\sum_{j=1}^{N_s}\left\Vert \alpha_{j}\right\Vert _{p}\:,
\end{equation}
where typically $p=0$ (or its relaxed convex version with $p=1$) and ${\bf \left\Vert \boldsymbol{X}\right\Vert }_{\mathrm{F}}=\left(\textrm{trace}(\boldsymbol{X}^{T}\boldsymbol{X})\right)^{1/2}$
is the Frobenius norm. For more technical details about GMCA, we refer the interested reader to \citet{bobin-gmca-cmb}, where it is shown that sparsity, as used in GMCA, allows for a more precise estimation of the mixing matrix $\bf A$ and more robustness to noise than ICA-based techniques.

Prior astrophysical knowledge can also be easily introduced in GMCA~: this includes the electromagnetic spectra of components such as CMB and SZ. 
BSS methods such as SMICA or CCA are intuitively well understood as they rely on second-order statistics astrophysicists customarily use in their everyday life~: correlations, covariance matrices, power and cross-spectra. 
More details can be found in \citet{bobin-gmca-cmb}.
Sparsity is not only sensitive to second order statistics but also to the higher-order statistics of the components. In \citet{bobin-gmca-cmb}, it has been shown on toy model simulations (all frequency channels at the same resolution and no spatial variation 
of the mixing matrix) that the sparsity criterion was more efficient than the SMICA criterion to separate the different components.

\subsection{Limitations of GMCA}
According to the mixture model underlying GMCA, all the observations are assumed to have the same resolution. However, in most CMB experiments, this assumption does not hold true~: the WMAP ({\it resp.} Planck) Full Width at Half Maximum (FWHM) varies by a factor of about 5 ({\it resp.} 7) between the best resolution and the worse resolution.\\
Assuming that the beam is invariant across the sky, the linear mixture model should be substituted with the following~:
$$
\forall i =1,\cdots,M; \quad x_i = b_i \star \left( \sum_{j=1}^n a_{ij} s_j \right) + n_i
$$
where $b_i$ stands for the impulse response function of the PSF at channel $i$. This model can be expressed more globally by introducing a global multichannel convolution operator $\mathcal{B}$ defined so that its restriction to the channel $i$ is equal to a convolution with $b_i$~:
\begin{equation}
\label{eq:beameq}
{\bf X} = \mathcal{B}\left( {\bf AS} \right) + {\bf N}
\end{equation}
To our knowledge, only blind source separation methods performed in spherical harmonics carefully account for the full mixture model in Equation~\ref{eq:beameq} where such a model can naturally be simplified (\textit{i.e.} the convolution operator is diagonalized by the spherical harmonics as in Fourier space for data sampled on regular Euclidean grids). However this holds true as long as the beam is assumed to be isotropic and invariant across the sky. Extending the data modeling to anisotropic and/or space varying beams can only be made in the pixel domain.\\ \\

The standard version of GMCA does not model for the different resolutions of the data. For the sake of simplicity, the effect of the beam was neglected during the source separation process. The estimation of the mixing matrix with GMCA was performed directly on the data assuming that the linear mixture model is valid. In this setting, the CMB map is evaluated by applying the Moore pseudo-inverse of the estimation mixing matrix to the raw data~: 
$$
s = [\boldsymbol{A}^+]_{1} \boldsymbol{X}
$$
where $\boldsymbol{A}^+ = (\boldsymbol{A}^T \boldsymbol{A})^{-1} \boldsymbol{A}^T$ and the notation $[\boldsymbol{Y}]_i$ stands for the $i$-th row of the matrix $Y$. Note that, by convention, matrix elements related to the CMB map have index $1$. Neglecting the beam effect implies that the CMB map estimated by GMCA is biased. Hopefully this bias can be computed explicitly by observing that, for each $(\ell,m)$ in the spherical harmonics domain~:
$$
s_{\ell,m} = \sum_{i=1}^{M} [\boldsymbol{A}^+]_{1,i} [\boldsymbol{A}]_{i,1} b_{i,\ell} s^\star_{\ell,m} + r
$$
where $r$ is a residual term that models for noise and foregrounds contaminating the estimated CMB. The variable $s^\star_{\ell,m} $ stands for the {\it true} bias-free CMB in spherical harmonics. By neglecting the residual term, the beam-induced bias can be regarded as a filter in spherical harmonics $\mathcal{B}_\ell$~:
\begin{eqnarray}
s_{\ell,m} & \simeq & \left[\sum_{j=1}^M [\boldsymbol{A}^+]_{1,j} [\boldsymbol{A}]_{j,1} b_{j,\ell} \right] s^\star_{\ell,m}  \\
& \simeq & \mathcal{B}_{\ell} s^\star_{\ell,m} 
\end{eqnarray}
where $\mathcal{B}_{\ell} = \sum_{j=1}^M [{\bf A}^+]_{1,j} [{\bf A}]_{j,1} b_{j,\ell}$. 
The final estimate is computed by inverting the above filter.\\ \\

Neglecting the beam effect in the component separation process has however two major drawbacks~: 
\begin{itemize}
\item{\bf Discrepancy with the linear mixture model~:} the linear mixture model does not hold when the data do not share the same resolution. This leads very likely to a non-optimal parameter estimation process. According to the linear mixture model, the contribution of the CMB in the data is the same across frequencies for each $(\ell,m)$~: this contribution is given by the electromagnetic spectrum of the CMB. This is no more true when the resolution varies from one observation to another~: this contribution now varies across $\ell$. This means that carrying out component separation from the raw data without taking care of the beam effect should lead to an inaccurate unmixing procedure. The dominant energetic content of the components is mainly concentrated at low frequencies where the beams do not differ too much from each other. At these frequencies, the linear mixture model is a good first order approximation which may explain the seemingly good performances of GMCA in \citet{Leach_08}. However, performances at smaller scales should be enhanced by correctly modeling the beam in the separation process.\\
\item{\bf Noise:~} Following the previous argument, the computation of the mixing matrix in GMCA is mainly driven by the low frequency content of the components. However the signal-to-noise ratio of the observations highly depends on their resolution; low resolution observations typically have a low SNR at high spatial frequencies. This entails that estimating the mixing parameters from the low frequency content for the data do not carefully account for the noise contamination at smaller scales; this is likely to lead to low SNR CMB estimates at high $\ell$.\\
\end{itemize}
Like most component separation methods used so far, GMCA explicitly assumes that the mixing matrix does not vary across pixels. This is a strong limitation as it is clearly not suited to capture the expected emissivity variation of galactic foregrounds across the sky.\\
We show in the next two sections how GMCA can be modified to solve these problems.

\section{GMCA and frequency map resolutions}

\subsection{Component separation from heterogenous data}
\label{sec:beams}

Accounting for the heterogeneity of the data in the separation process can be carried out in two different ways~: the most straightforward technique would consist in adapting GMCA by substituting the data fidelity term in Equation~\ref{eq:GMCA2} with the more rigorous expression~: $\left \| {\bf X} - \mathcal{B}\left({\bf AS} \right) \right \|_{F,{\bf \Sigma_N}}^2$ where $\mathcal{B}$ represents the beam effect. If this approach has been explored with some success in a different imaging context \citep{SPIEGMCA09}, its high computational cost makes it hard to apply to large-scale CMB data.\\
The second approach would rather consist in finding ways to apply GMCA to data that share the same resolution. 
This could be obtained by first converting all the observations to the common desired resolution ({\it e.g.} $5$ arcmin). 
However, the beams at low frequency vanish much faster than at higher frequency; 
this entails that a brute-force deconvolution of the low frequency channels will yield a large amplification of the noise. In practice, such a deconvolution is also prohibited by numerical issues.
It is important to underline that low resolution channels will mainly add noise to the final estimate at high spatial frequencies (or equivalently high $\ell$ multipoles in the spherical harmonics domain). 
Therefore low resolution observations will be most informative at low frequencies and should be avoided for the reconstruction of the high frequency content of the CMB map. In the next paragraph, we introduce an elegant strategy to extend GMCA so as to cope with data involving frequency-dependent beams.

\subsection{Multiscale GMCA}
A solution to this problem is to adapt the wavelet decomposition for each channel such that the wavelet coefficients ${w_{1}^{(\mu)}, ..., w_{M}^{(\mu)}}$ of the $M$ available channels at scale $\mu$ do have exactly the same resolution.
This can be easily obtained by choosing a specific wavelet function for each channel $i$ ($i=1,\cdots,M$) such that:
\begin{equation}
{\Psi}^{(\mu)}_{i}  (\ell) =   \frac{ b_{target}(\ell)} {b_i(\ell)}   {\psi}^{(\mu)}  (\ell) 
\end{equation}
where $b_i$ is the beam of the $i$-th channel, $b_{target}$ is the Gaussian beam related to the targeted resolution, and ${\psi}_\mu $ is the standard wavelet function at scale $\mu$.
This approach is closely related to the wavelet-based deconvolution techniques introduced in \citet{rest:donoho95b,rest:kalifa00} and called, in the statistical literature, wavelet-vaguelette decomposition. 
According to this formalism, the wavelet-vaguelette decomposition of a function $f$ is defined by
\begin{equation}
b_{target} \star f = \sum_{\mu=1}^{N_\mu}  \sum_{k=1}^{N_p}   \left< b_i \star f , \Psi^{(\mu)}_{k} \right> \psi^{(\mu)}_{k} 
\end{equation}
where $N_\mu$ is the number of wavelet scales and $N_p$ is the number of pixels in the map.\\

In this paper, we propose extending these ideas to the case of component separation. Replacing the data matrix ${\bf X}$ by the matrix of wavelet coefficients $\bf W$ (${\bf W}_{i,*} = w_i=  \left< x_i, \Psi^{(\mu)}_{i,k} \right>$), and
applying the pseudo inverse of ${\bf A}$ to ${\bf W}$, we get ${\bf Z} = \boldsymbol{A}^+ {\bf W}$. The quantity $z_{j,k}^{(\mu)}$  is now related to the wavelet coefficients of the $j$-th source in the $\mu$-th wavelet scale at location $k$. The estimated sources can then be estimated via the following reconstruction formula :
\begin{equation}
\tilde s_j =  \sum_{\mu=1}^{N_\mu}  \sum_{k=1}^{N_p}  z_{j,k}^{(\mu)}  \psi^{(\mu)}_{k}
\end{equation}
where the notation $\tilde s_j$ denotes the estimate of the $j$-th source.
The reconstruction formula cannot be implemented in practice, because the wavelet coefficients of the $i$-th channel $w^{(\mu)}_{i}$ at scale $\mu$ cannot be calculated when the fraction 
$\frac{ b_{target}(\ell) } {b_i(\ell)}  {\psi}^{(\mu)}  (\ell) $ is undefined or numerically unstable. It means that at each wavelet scale $\mu$, only the channels with non-vanishing beams can be used. 
At the largest wavelet scale, all channels are active, while at the finest wavelet scale, only few channels are active. \\ \\
We note  $C_{\mu}$ the number of channels available for a given resolution level $\mu$, and $N_{j}^{(\mu)}$ the number of wavelet scales that can be used for a given source $j$. This implies introducing a mixing matrix ${\bf A}^{(\mu)}$ per wavelet scale; this mixing matrix will be evaluated from the $C_\mu$ channels available at scale $\mu$.
The size of these matrices and the number of sources (which is limited to the number of channels) are varying with $\mu$. For each wavelet resolution level $\mu$, we now have a solution $\tilde s_j^{(\mu)}$~:
\begin{equation}
\tilde s^{(\mu)}_j =   \sum_{k=1}^{N_p}   z^{(\mu)}_{j,k}    \psi_{k}^{(\mu)}
\end{equation}
where  ${\bf Z}^{(\mu)} = {{\bf A}^{(\mu)}}^+  {{\bf W}^{(\mu)}}$, and ${w}^{(\mu)}_i =   \left< x_i, \Psi^{(\mu)}_{i,k} \right>$, with $i = 1 ,\cdots, C_{\mu}$, and $\mu=1,\cdots,N_{\mu}$.           

The final solution $s_j$ for the $j$-th source is obtained by a simple wavelet reconstruction~:
\begin{eqnarray}
\tilde s_j & =  & \sum_{\mu=1}^{N_j^{(\mu)}}  \sum_{k=1}^{N_p}   \left<  {\tilde s}^{(\mu)}_j,  \psi_{k}^{(\mu)} \right>   \psi_{k}^{(\mu)}  
\label{eq_aggreg_wave}
\end{eqnarray}
Multiscale GMCA (mGMCA) is similar to a harmonic space method, where we consider one mixing matrix per wavelet band (or frequency band), but on the contrary of SMICA, the mixing matrix is calculated from high order statistics of wavelet coefficients. Like SMICA, mGMCA can properly take into account the resolution of the different channels but with the paramount advantage that it does not make any assumption on the stationarity of the sources. 
 
\subsection{Practical Implementation}
Though the wavelet-vaguelette source separation method seems a complicated procedure, it can be largely simplified. Indeed, thanks to the linearity of both the beam convolution operator and the wavelet operator, the matrix ${\bf W}^{(\mu)}$  relative to the active channels at resolution level $\mu$ can be computed by putting the active maps at the same resolution (which depends on $\mu$) and then compute the same wavelet decomposition on each of them. Once the matrix ${\bf W}^{(\mu)}$ is obtained, one can run the GMCA algorithm to get the mixing matrices ${{\bf A}^{(\mu)}}$.
This can be repeated for each resolution level $\mu$. This way, a CMB map is evaluated for each resolution level, with a number of active channels decreasing with scales. 
The final solution is then derived by aggregating these solutions in the wavelet space as in Equation~\ref{eq_aggreg_wave}.\\
Another advantage of this approach is its ability to benefit from the structure of the Healpix format~: different values for the parameter {\it nside} can be chosen depending on the resolution level, which speeds up the computation time.\\
As an illustration, we give a possible parameterization of mGMCA for Planck data. In this case, we have nine channels from 30 to 857 GHz, with a resolution which goes roughly from 33 arcmin to 5 arcmin. We therefore have considered five resolution levels, with a number of active channels varying from 
9 to 5 (see Table~\ref{tab_resol_planck}).  
\begin{table}
\begin{center}
\vspace{0.1in}
\begin{tabular}{|c|c|c|}
\hline
Observations used & Scale & ${C_{\mu}}$ \\
\hline
\hline
30 to 857 GHz & 33 arcmin & 9 \\
\hline
44 to 857 GHz & 24 arcmin & 8 \\
\hline
70 to 857 GHz & 14 arcmin & 7  \\
\hline
100 to 857 GHz & 10 arcmin & 6  \\
\hline
143 to 857 GHz & 5 arcmin & 5  \\
\hline
\end{tabular}
\vspace{0.1in}
\end{center}
\caption{Example of resolution levels to use in mGMCA, with the number of active channels per resolution level.}
\label{tab_resol_planck}
\end{table}

However, the underlying modeling of mGMCA does not allow for a precise separation of components with spectral variations. Next section will show that the spatial variation of the matrix can also be taken into account using a partitioning of the wavelet scales.


\section{GMCA and spatially variant mixing matrix}
\label{sec:lgmca}

\subsection{A refined modeling to get closer to astrophysics}
\label{sec:astropriors}

As emphasized in the introduction, the complexity of CMB data makes it very hard to fully model for all the physical phenomena with a simple linear mixture model. The linear mixture model used so far in most component separation methods assumes that~: i) the number of components is limited to the total number of observations, ii) explicitly assumes that the emissivity of the component is {\it space invariant} ({\it i.e.} the mixing matrix does not vary from one pixel to another). Unfortunately, these assumptions are not verified by common CMB data. The components one observes between $30$GHz and $857$GHz include the CMB, Sunyaev Zel'Dovich effect, free-free, synchrotron, CIB, anomalous dust and dust emissions as well as IR and radio sources the number of which largely exceeds the number of observations provided by Planck.\\
From the current knowledge in astrophysics, some of these components can be approximated quite accurately with space-invariant emissivity; this is the case for the CMB, SZ effect, free-free and synchrotron emission.\\
The dust emissions provide a very important, if not dominant, contribution in high frequency channels. These channels have the best spatial resolution (\textit{typ.} $5$ to $10$ arcmin); modeling for such a component should thus be of utmost importance for a high-resolution estimate of the CMB map. However modeling dust emission is a strenuous problem. Indeed, contrary to well-characterized foreground emissions such as free-free or synchrotron, most component separation models do not provide a satisfactory modeling of this contribution. As an example, the gray body dust model, known as being one of the most accurate dust model, involves two parameters~: a spectral index and temperature varying across pixels.\\
These remarks imply that the linear mixture model does not allow for enough degrees of freedom to fully capture the complexity of CMB data in the frequency range observed by most CMB surveys like Planck. In the following, we will focus on extending this model to account for spatially-variant spectra.



\subsection{Multiscale local mixture model}
\label{sec:lmm}

The global mixture model used by most component separation methods do not allow for enough degrees of freedom to capture naturally scale/space-dependent astrophysical phenomena. Scale-dependancy of the analysis naturally arises from the mGMCA formalism, localization requires decomposing each wavelet scale into patches. It is worth mentioning that localizing the estimation of the CMB has also been proposed within the ILC framework \citep{NeedletILC} to analyze WMAP map data. N-ILC consists in~: i) decomposing each wavelet scale into patches with a scale-dependent size, ii) performing ILC on each patch independently. However, the spectral behavior of many components is not expected to change dramatically from one patch to its neighbor; such an independent processing each on patch may not be optimal.\\ 

\begin{figure}[htb]
\centerline{\includegraphics[scale=0.4]{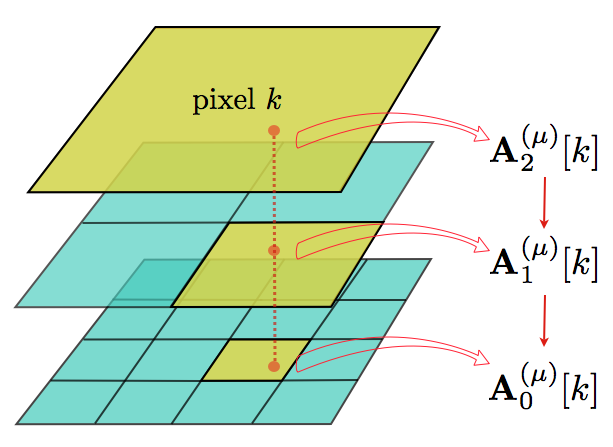}}
\caption{Multichannel quadtree decomposition. This figure displays the tree-based decomposition of the multichannel data ${\bf W}^{(\mu)}$. A sequence of mixing matrices are estimated on patches with dyadically decreasing size. Large size patches are used as a warm startup for the estimation process on smaller size patches.} \label{fig:quadt}
\end{figure}

For this purpose, we extend the mixture model to a {\bf multiscale local} mixture model. In such modeling, each location of the sky in each wavelet scale is analyzed several times with different patch sizes which allows to locally select the best parameters.\\ 
Before going any further, let us first recall some useful notations~: if $\bf X$ denotes the data, we will denote by ${\bf W}^{(\mu)}$ the matrix composed of the $\mu$-th wavelet scale of the data $\bf X$. In what follows, the indexing ${\bf W}^{(\mu)}[k]$ will denote the square patch of size $p$ centered about pixel $k$. Following the multiscale local model, a patch-based representation of the data at scale $\mu$ and location $k$ is modeled as follows~:
\begin{equation}
\label{eq:mlml}
{\bf W}^{(\mu)}[k] = {\bf A}^{(\mu)}[k] {\bf S}^{(\mu)} [k] + {\bf N}^{(\mu)}[k]
\end{equation}
A direct extension of mGMCA to solve this problem would simply amount to applying GMCA independently to each patch at location $k$ and each scale $\mu$. However, this approach would suffer from certain drawbacks assuming that some "optimal'' patch size at each scale $\mu$ is known and fixed in advance. However, fixing \textit{a priori} the patch size is a very strong constraint~: the appropriate patch size should be space-dependent as well and may vary from one region to another.\\
This suggests that a trade-off should be made between small/large patches which would balance between statistical consistency (large patches) and adaptivity (small patches). This indicates that the choice of the patch-size should be adaptive and dependent on the local content of the data.
Inspired by best basis techniques in multiscale signal analysis, an elegant way to alleviate this pitfall is to perform GMCA at each wavelet scale $\mu$ with various patch sizes in a quadtree decomposition. In a nutshell, GMCA is first performed on the full field to obtain a first estimator of the mixing matrix denoted by ${\bf A}_j^2[k]$ in Figure~\ref{fig:quadt}. The field is further decomposed into $4$ identical non-overlapping patches on which GMCA is applied to provide a set of mixing matrices denoted ${\bf A}_j^1[k]$ in the figure. The process is iterated until the patch is equal to the desired patch size $p_\mu$. The number of analysis levels is equal to $L_\mu$.
Interestingly, the same area of the sky being analyzed at different mixture scales (\textit{i.e.} patch sizes), it makes it possible to choose afterwards the "optimal'' patch size from the different estimates obtained for $l = 0,\cdots,L_\mu$. It therefore allows for more degrees of freedom to select an adapted patch size at each location. An exhaustive description of the parameter selection strategy is given in Appendix~\ref{sec:localestim}.\\

\subsection{The L-GMCA algorithm}
\label{sec:algo}

The Local GMCA (L-GMCA) algorithm can be described as follows~:

\begin{center}
\vspace{0.15in}
\centering
\begin{tabular}{|c|} \hline
\begin{minipage}[hbt]{0.95\linewidth}
\vspace{0.025in}
\footnotesize{

\textsf{1. Compute the wavelet decomposition of the data.} \\

\textsf{2. For each wavelet scale $\mu = 1,\cdots, N_\mu$~:} \\

\hspace{0.2in} \textsf{2.1 Put the $C_{\mu}$ active channels to the same targeted resolution} \\

\hspace{0.2in} \textsf{2.2 For each local $k$~:} \\

\hspace{0.4in} \textsf{2.2.1 - Compute a set of mixing matrices at different patch sizes ($L_\mu$ levels of analysis)} \\

\hspace{0.4in} \textsf{2.2.2 - Select the optimal mixing matrix} \\

\hspace{0.4in} \textsf{2.2.3 - Compute the CMB map estimate at scale $\mu$ and location $k$.} \\

\textsf{3. Reconstruct the CMB map following Equation~\ref{eq_aggreg_wave}}}  \\

\vspace{0.05in}
\end{minipage}
\\\hline
\end{tabular}
\vspace{0.15in}
\end{center}

Once the mixing matrices are estimated with L-GMCA (steps $2.2.1$ and $2.2.2$ of the algorithm), estimating the CMB map requires to perform both a wavelet-vaguelette decomposition and a weighting of the wavelet coefficient of the multifrequency data at each wavelet scale according to the pseudo-inverse values of the estimated local mixing matrix. This sequence of operations is particularly computer intensive for high resolution data ({\it e.g.} $Npix \simeq 50$ million pixels for Planck data for each of the 9 frequency maps), requiring for each resolution level and each channel two back and forth spherical harmonic transform (of asymptotic complexity $Npix^{3/2}$) followed by two wavelet transform for each HEALPix face (of asymptotic complexity $Npix \log(Npix)$), and a weighting of each coefficient map at each scale and each resolution level (of asymptotic complexity $Npix$). This should be multiplied by the number of Monte-Carlo simulations.
We developed a C++ parallelized program using OpenMP. As an illustration of how crucial computer speed might be, on a shared-memory multiprocessor system containing 8 Six-Core AMD Opteron(tm) running at 2.4 GHz, recovering the CMB map from one Monte-Carlo simulation takes about 9 minutes when 24 processors are used.


\section{Experiments}
\label{sec:results}

\subsection{Simulations and comparisons}

In early 2013, the finest - low noise, high resolution - CMB data will be Planck data. Therefore, in this section, we propose evaluating the performances of the proposed approach on simulated Planck data. To the best of our knowledge, the only publicly available full-sky simulated Planck data are the dataset the used to evaluate preliminary performances of various CMB map estimation techniques in 2008. This pre-launch study has been published in \citet{Leach_08}. In the sequel numerical experiments will be performed using exactly the same Planck simulated dataset. Before going deeper into the description of the numerical results, the two following paragraphs clarify the astrophysical content of the simulation as well as the evaluation protocol we opted for.

\paragraph{Some words about the simulations~:} 

The Planck-like dataset described in \citet{Leach_08} has been simulated using an early version of the Planck Sky Model (PSM) developed by J. Delabrouille and collaborators\footnote{For more details about the PSM, we invite the reader to visit the PSM website~: {\it http://www.apc.univ-paris7.fr/~delabrou/PSM/psm.html}.}. In 2008, the PSM included most astrophysical signals and foregrounds as well as simulated instrumental noise and beams. In details, the simulations were obtained as follows~:
\begin{itemize}

\item{\bf Frequency channels~:} the simulated data are comprised of the $9$ LFI and HFI channels at frequency $33,44,70,100,143,217,353,545,857$ GHz. The frequency-dependent beams are perfect isotropic Gaussian PSF with FWHM ranging from $5$ arcmin at $217,353,545,857$ GHz to $33$ arcmin at $33$GHz. Real data differ from the model as the scanning strategy adopted by full-sky surveys such as WMAP or Planck is likely to produce non-isotropic and spatially varying beams which are more elongated along the scanning direction.\\

\item{\bf Instrumental noise~:} in full-sky surveys, the sky coverage is generally not uniform~: some areas are more often observed than others. To some extent, this entails that the instrumental noise variance is not homogenous as well. In statistical signal processing, such kind of statistical process is better known as heteroscedastic noise. A second effect of the scanning strategy is the correlation of the noise along the scanning direction. In the sequel, simulations account for the non-homogeneity of the noise but noise samples are assumed to be uncorrelated. Either for WMAP or Planck, the noise statistics ({\it i.e.} noise variance map for each frequency channel) are assumed to be known accurately.\\

\item{\bf Cosmological Microwave Background~:} the CMB map is a correlated random Gaussian realization entirely characterized by its power spectrum. In the simulations, the CMB power spectrum is identical to that of WMAP. Higher order multipoles are based on a WMAP best-fit power spectrum at high $\ell$. Note that the simulated CMB is Gaussian; no non-gaussianity ({\it e.g.} lensing, ISW, $\mbox{f}_{NL}$) has been added. This will allow for non-gaussianity tests under the null assumption in the sequel.\\

\item{\bf Dust emissions~:}  the galactic dust emissions is composed of two distinct dust emissions~: thermal dust and spinning dust ({\it a.k.a.} anomalous microwave emission). Thermal dust is modeled via the Finkbeiner model \citep{Fink99} which assumed two hot/cold dust populations contribute to the signal in each pixel. The emission law of thermal dust varies across the sky.\\

\item{\bf Synchrotron emission~:} The synchrotron emission, as simulated by the PSM, is an extrapolation of the Haslam $408$MHz map \citep{Haslam:1982dk}. The emission law of the synchrotron emission is an exact power law with a spatially varying spectral index.\\

\item{\bf Free-free emission~:} the spatial geometry of free-free emission is inspired by the H$\alpha$ map built from the SHASSA and WHAM surveys. The emission law is a perfect power law with a fixed spectral index.\\

\item{\bf Point sources~:} infrared and radio sources were added based on existing catalogues at that time. In the following, the brightest point sources will be masked prior to the results evaluation.\\

\item{\bf Sunyaev Zel'Dovich effect~:} thermal SZ is modeled in the simulations.\\

\end{itemize}

Small scales of the foreground maps were extrapolated to the Planck resolution. More details about the simulations can be found in \citet{Leach_08}.

\paragraph{Comparison protocol~:} 

In 2008, most of the estimated CMB maps used for comparisons were quite heterogenous~: they did not necessarily share the same beams. As a consequence, precise and quantitative evaluations were very hard to carry out. 
However, with the exception of GMCA\footnote{GMCA is available at this address~: {\it http://jstarck.free.fr/isap.html}}, none of the codes used to estimate CMB maps in \citet{Leach_08} is publicly available.\\
The major objective of this paper is to evaluate the impact of the local and multi scale mixture model as well as the sparsity-based component separation technique we introduced in this paper.\\
Since WMAP, ILC and its extensions \citep{NeedletILC,GenILC} have taken the lion's share for full-sky CMB data analysis; it is therefore very popular in the astrophysics community. Like most component separation techniques in cosmology, ILC relies on second order statistics (more precisely, $\chi^2$ minimization) to estimate the CMB map. For this reason, we first chose to compare two different separation methods~: GMCA (based on sparsity) and ILC (based on second order statistics). Furthermore, we believe that one of the contributions of this paper is the use of the local modeling in the wavelet domain; we will also evaluate the performances of methods along with a global mixture model (the corresponding methods will be ILC and GMCA) as well as with the local mixture model in wavelets (L-GMCA).\\

\begin{table*}
\begin{center}
\vspace{0.1in}
\begin{tabular}{|c|c|c|c|c|}
\hline
Band $\mu$ & \# common resolution & \# sources & \# quad-tree levels : $L_\mu$ & Nominal patch size : $p_\mu$ \\
\hline
\hline
1  &  33 arcmin & 9 & 3 & 64 \\
\hline
2 &  24 arcmin  &  8  & 3 & 64 \\
\hline
3 &  14 arcmin  & 7  & 3 & 32 \\
\hline
4  &  10 arcmin & 6 & 3 & 32 \\
\hline
5  &  5 arcmin & 5 &  3 & 16 \\
\hline
\end{tabular}
\vspace{0.1in}
\label{tab:params}
\caption{This table details the parameters used in local and multiscale sky model. The number of wavelet-vaguelette scales is $N_\mu= 5 $. }
\end{center}
\end{table*}

The local and multiscale model requires the definition of $4$ parameters~: i) the number of sources is set to be equal to the number of channels, ii) the number of wavelet scales, iii) the number of quad-tree decomposition levels $L_\mu$ and  iv) the nominal patch size $p_\mu$. All these parameters are given in Table~\ref{tab:params}. There is no automatic strategy to optimally select these parameters. We would like to notice that a non-exhaustive series of experiments; the particular choice of Table~\ref{tab:params} turned out to be provide lower power spectrum bias, especially at high $\ell$. NILC (Needlet-ILC) is also performed \cite{NeedletILC,NILC12}.\\
In the sequel, a $92$\% common mask, which preserves a very large part of the sky, has been defined as the union of a small galactic mask and point sources mask which has been derived from the brightest point sources.

\subsubsection{CMB map estimation}
\begin{figure*}[ht]
\centerline{\includegraphics[scale=0.3]{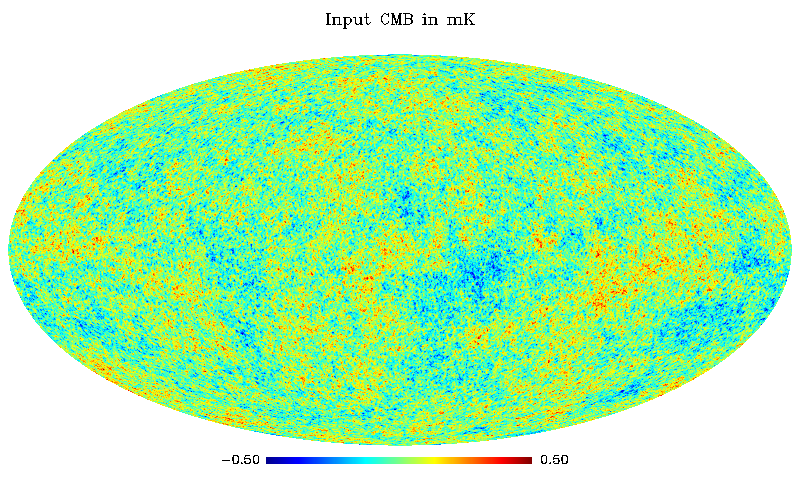}}
\begin{tabular}{cc}
\includegraphics[scale=0.3]{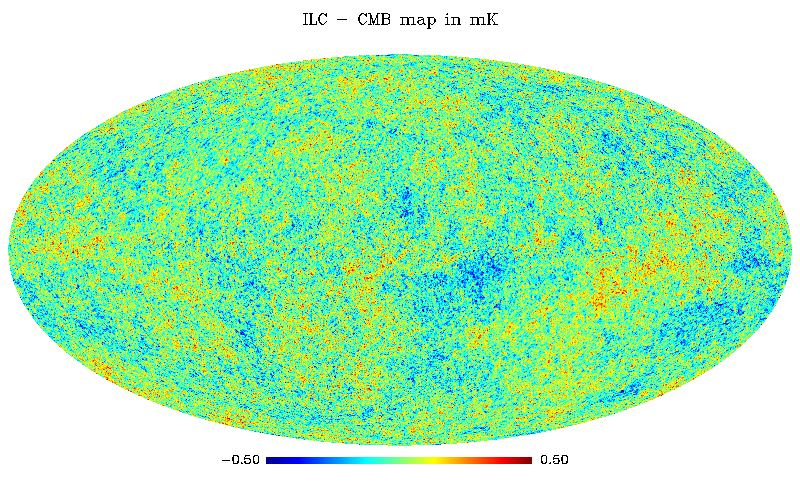} & \includegraphics[scale=0.3]{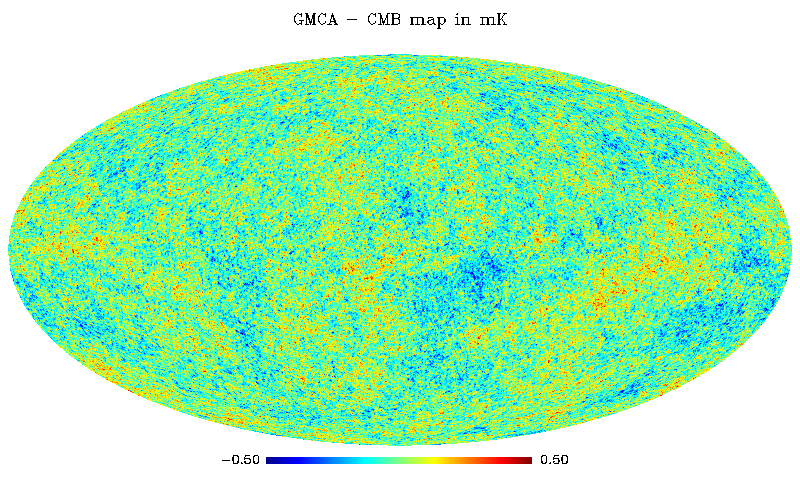} \\
\includegraphics[scale=0.3]{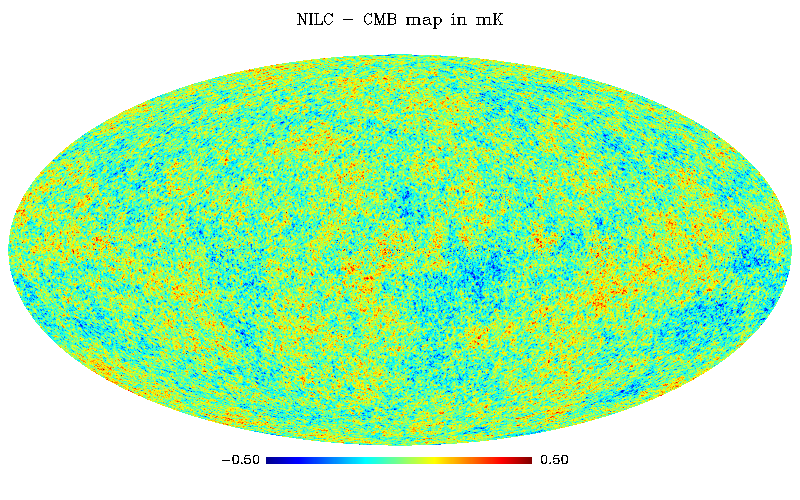} & \includegraphics[scale=0.3]{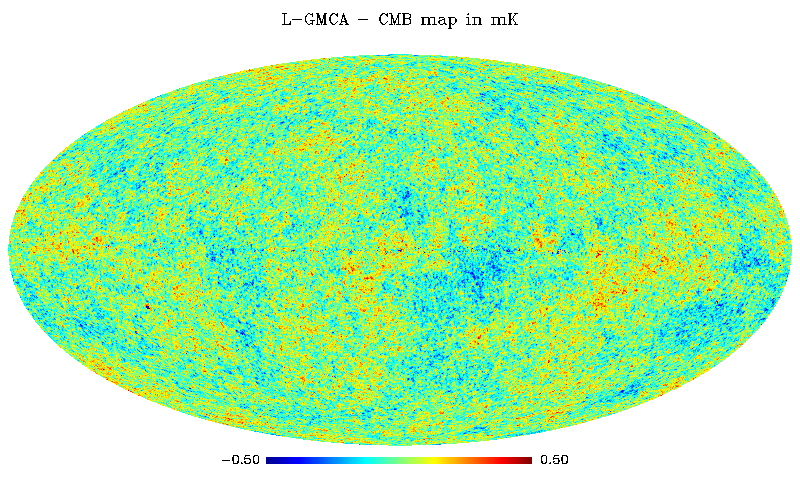}
\end{tabular}
\caption{Input (top) and estimated CMB maps in mK.}
\label{fig:cmb_maps}
\end{figure*}

\begin{figure*}[ht]
\begin{tabular}{cc}
\includegraphics[scale=0.3]{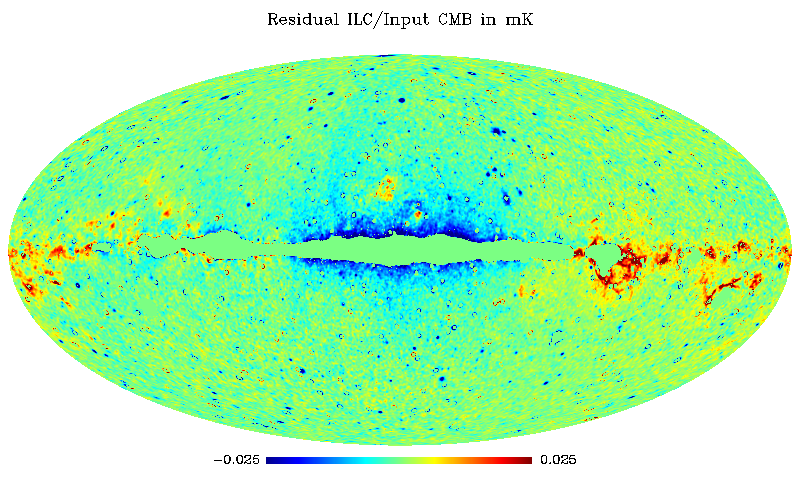} & \includegraphics[scale=0.3]{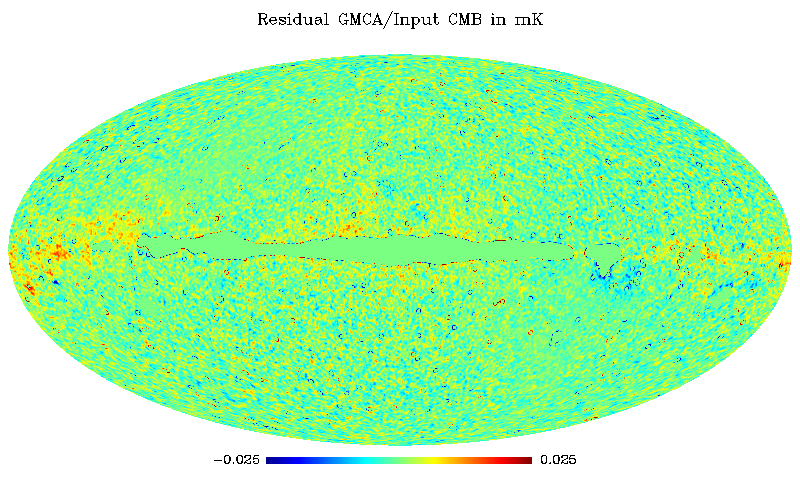} \\
\includegraphics[scale=0.3]{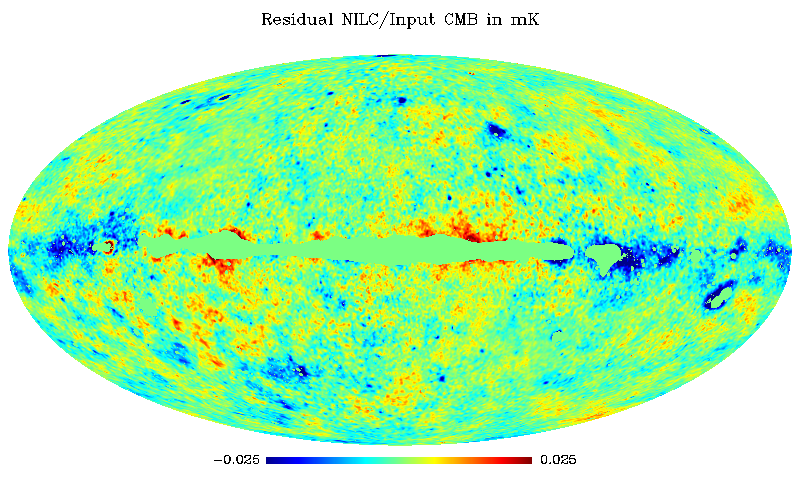} & \includegraphics[scale=0.3]{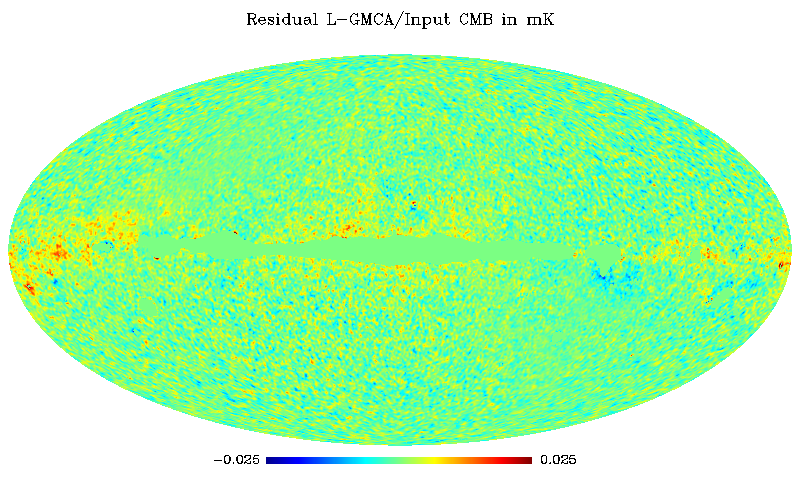}
\end{tabular}
\caption{Residual CMB maps. These maps are defined as the difference between the estimated maps and the input CMB map. Units in mK.}
\label{fig:residual_maps_45}
\end{figure*}

\begin{figure*}[ht]
\center \includegraphics[scale=0.2]{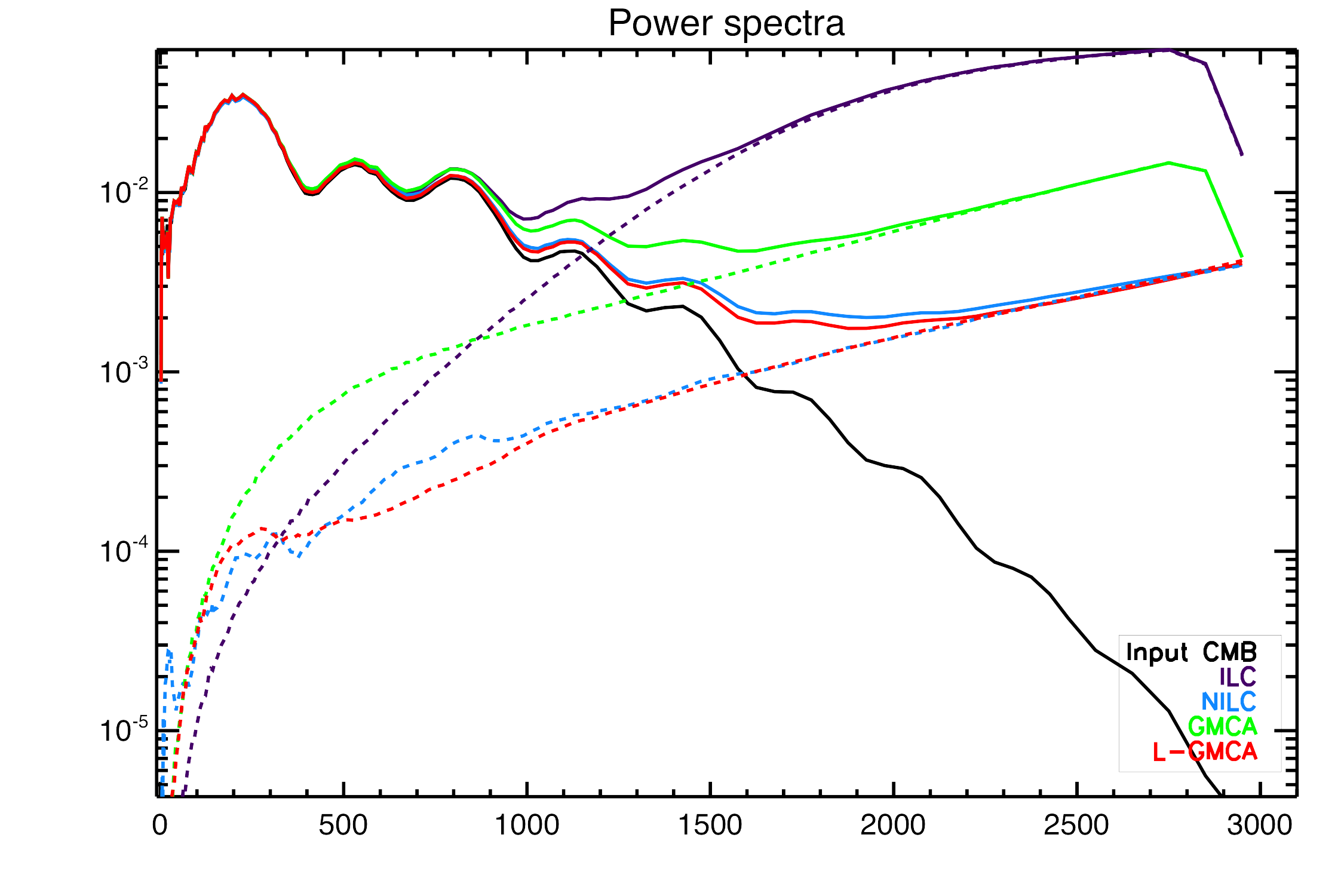} 
\caption{Power spectrum of the maps. The dotted line is power spectrum of the input CMB map at the resolution of $5$ arcmin. {\it Abscissa~:} spherical harmonics multipoles $\ell$. {\it Ordinate~:} Power spectra value - units in $\ell(\ell+1)C_{\ell}/(2\pi) \mbox{ mK}^2$}
\label{fig:ps_error0}
\end{figure*}

In this paragraph, we mainly focus on the quality of estimation of the full-sky CMB map. Figure~\ref{fig:cmb_maps} display the input CMB at a resolution of $5$ arcmin on top and the maps we obtained by performing ILC, GMCA, NILC and L-GMCA. More interestingly, Figure~\ref{fig:residual_maps_45} show the residual maps we defined as the difference between the estimated maps and the input CMB map at $5$ arcmin. These error maps have further been filtered at the resolution of $45$ arcmin to filter out noise contamination which dominates at high frequency; this allows to better unveil the low frequency foreground residuals which remain in the final estimates. The ILC residual map (top-left picture) clearly exhibits large-scale features which are reminiscent of the synchrotron emission (negative bulb centered about the galactic center) as well as free-free emissions relics (well-known positive free-free on the right of the residual map) and dust emission. The NILC residual show significant large scale "blobby" effects which may be related to the use of needlet filters used for the analysis rather than more space-localized wavelet filters. Visually, GMCA, and L-GMCA seem to have similar remaining foreground contamination which is evocative of dust emission.\\
Figure~\ref{fig:ps_error0} shows the power spectra of the CMB maps. These spectra seem very similar up to $\ell = 500$. At smaller scales, instrumental noise is the dominant signal. Dotted lines feature the noise power spectra. Global methods, and especially ILC, exhibit a large noise level at high $\ell$; these methods do not account for the beam variation across channels. Therefore, ILC and GMCA are mainly sensitive to the largest scales and do not carefully deal with noise contamination at smaller scales. Space/scale localized approaches like L-GMCA and NILC have lower noise contamination with only slight differences between them. It is remarkable that, if NILC is expected to exhibit low noise contamination ({\it i.e.} it relies on the second order statistics of the data), L-GMCA is also well designed to provide low noise contamination levels. 


%
%

\subsubsection{Power spectrum estimation}
Measuring the power spectrum of the CMB map is of utmost importance for cosmology; crucial information can be derived to constrain the cosmological parameters. Power spectrum estimation is generally a strenuous problem~: a precise astrophysical analysis of the CMB power spectrum has generally to be carried out to carefully evaluate its precise value and confidence intervals on these values. The common approach generally consists in estimated the CMB power spectrum from the frequency channels with very large masks to reduce the impact of galactic foregrounds. The use of a single channel allows for more handy studies of contaminants impacts ({\it e.g.} point sources, cosmological infrared background). However, the mandatory use of very large masks dramatically increases estimation errors at large-scales. The main advantage of source separation is its ability to clean large areas of the sky which makes possible the use of smaller masks. The final being a non trivial combination of the frequency channels, studying error propagation is generally more complex. Fortunately, the linearity of the proposed methods makes possible the use of Monte-Carlo simulations to further study how foregrounds contaminate the final CMB map.\\
We carried out comparisons of the CMB power-spectrum which can be computed from the CMB maps estimated by these $4$ methods. These experiments showed that the local/multiscale mixture model yields less bias but did not reveal significant differences between ILC-based and sparsity-based component separation techniques. These non-conclusive results can be explained by the low sensitivity of the power spectrum to slight foreground / non-gaussian contamination. Interestingly, the results published in \citep{Leach_08} were quite comparable at all scales thus also leading to non-conclusive comparisons at the level of the CMB power-spectrum.

\subsection{Foreground contamination}
\label{sec:crosscor}

\begin{figure}[h!]
\begin{tabular}{c}
\includegraphics[scale=0.2]{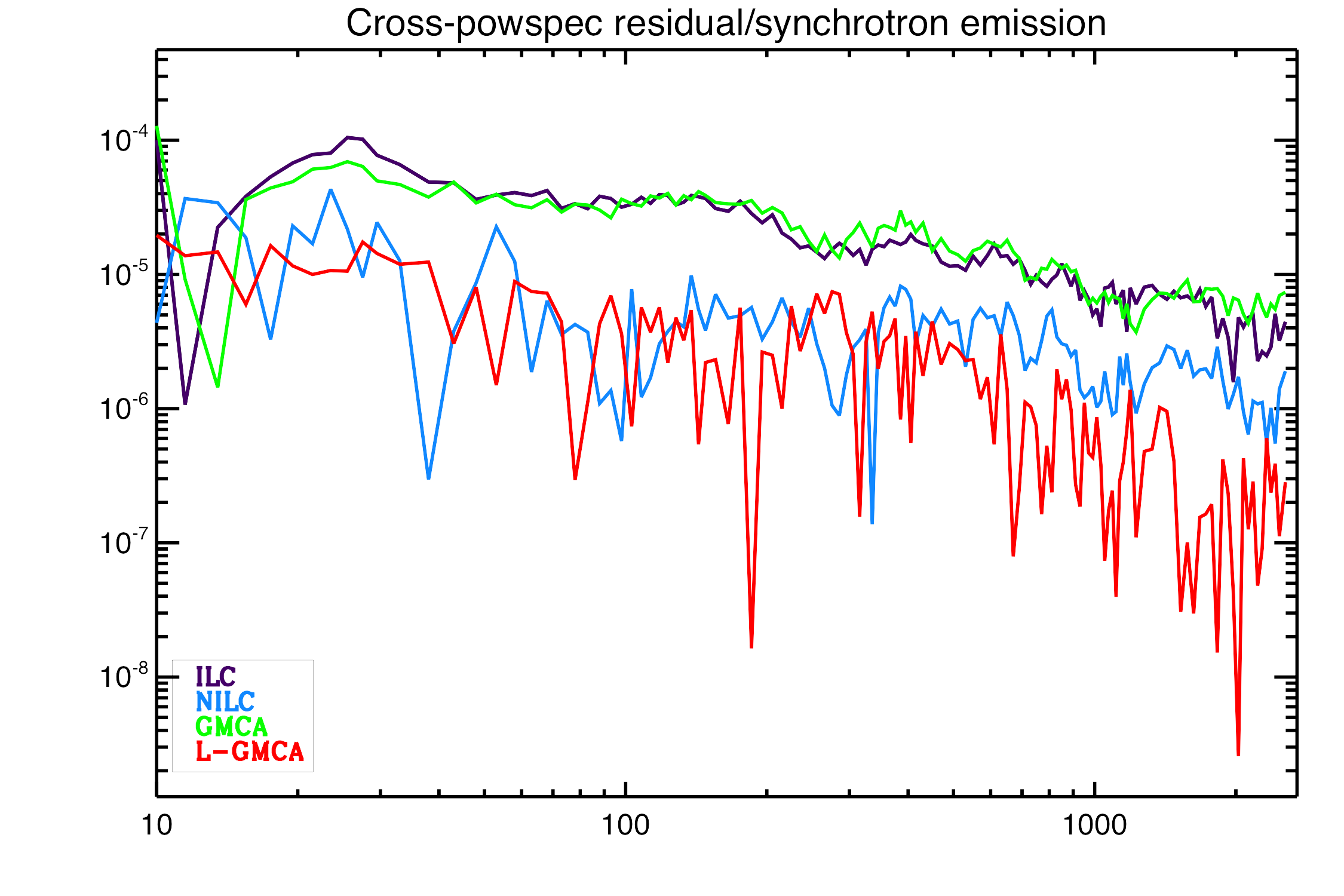} \\
 \includegraphics[scale=0.2]{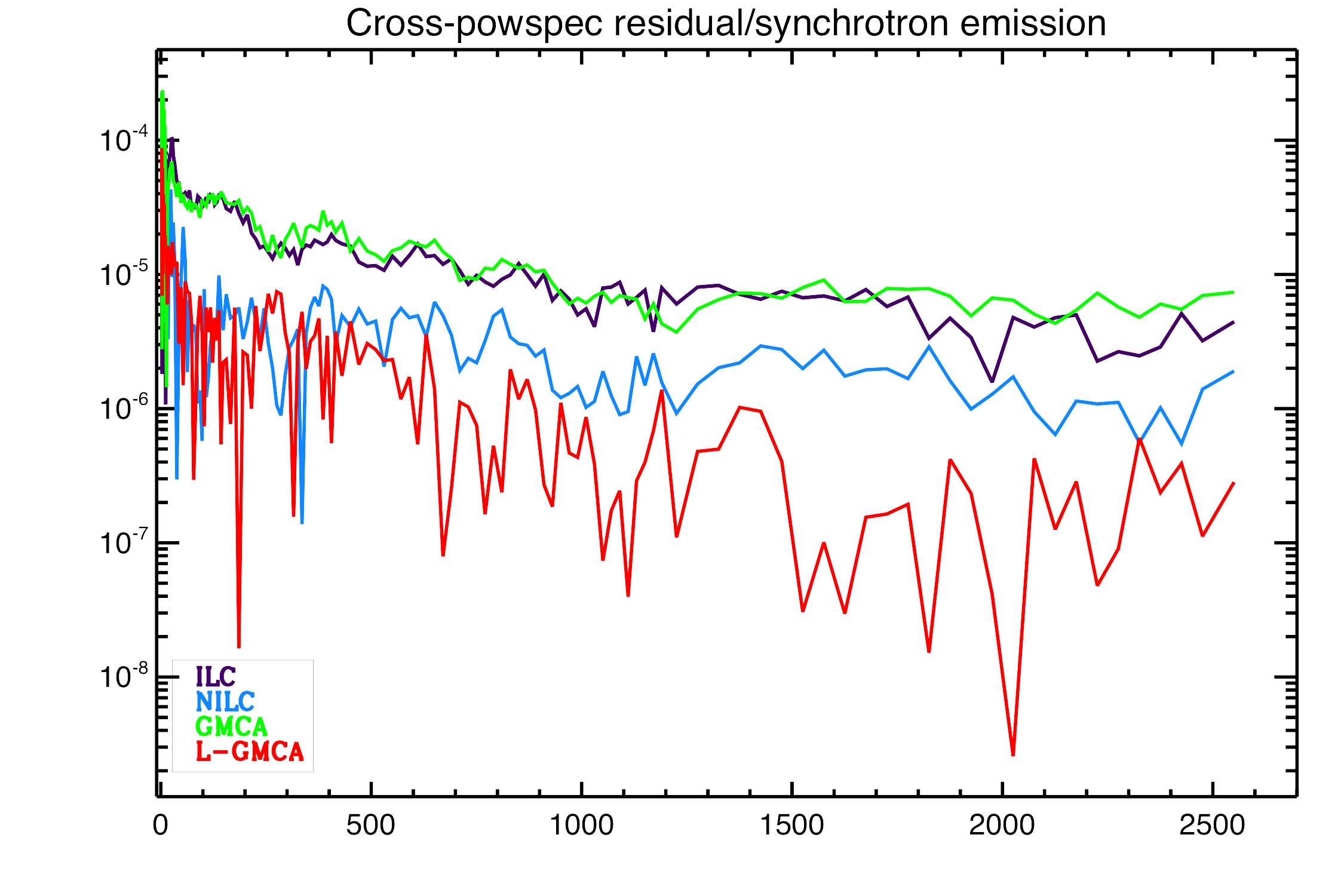} 
\end{tabular}
\caption{Cross-power spectra of the estimated CMB map with the synchrotron template.} 
\label{fig:sync_ct}
\end{figure}

\begin{figure}[h!]
\begin{tabular}{c}
\includegraphics[scale=0.2]{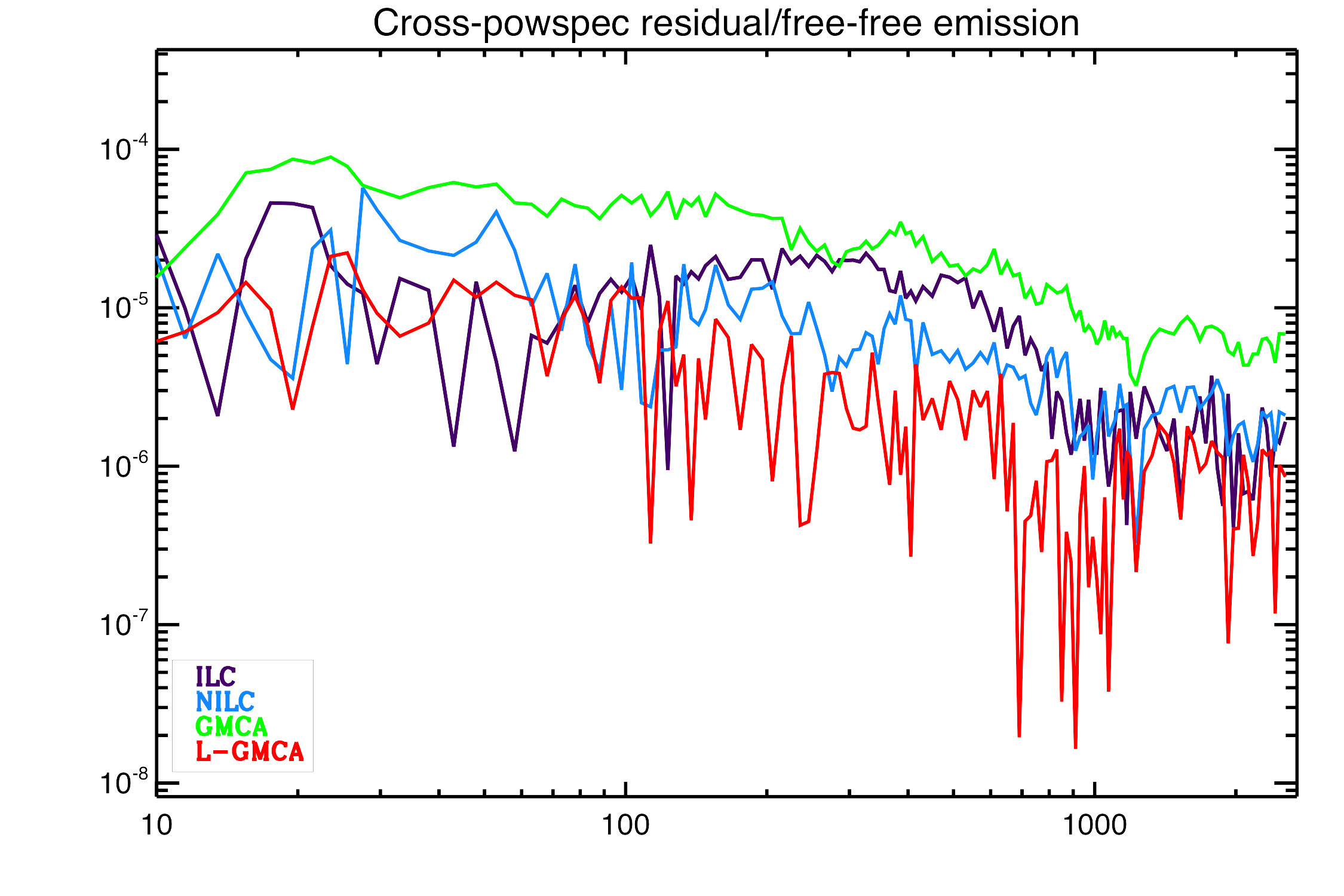} \\
 \includegraphics[scale=0.2]{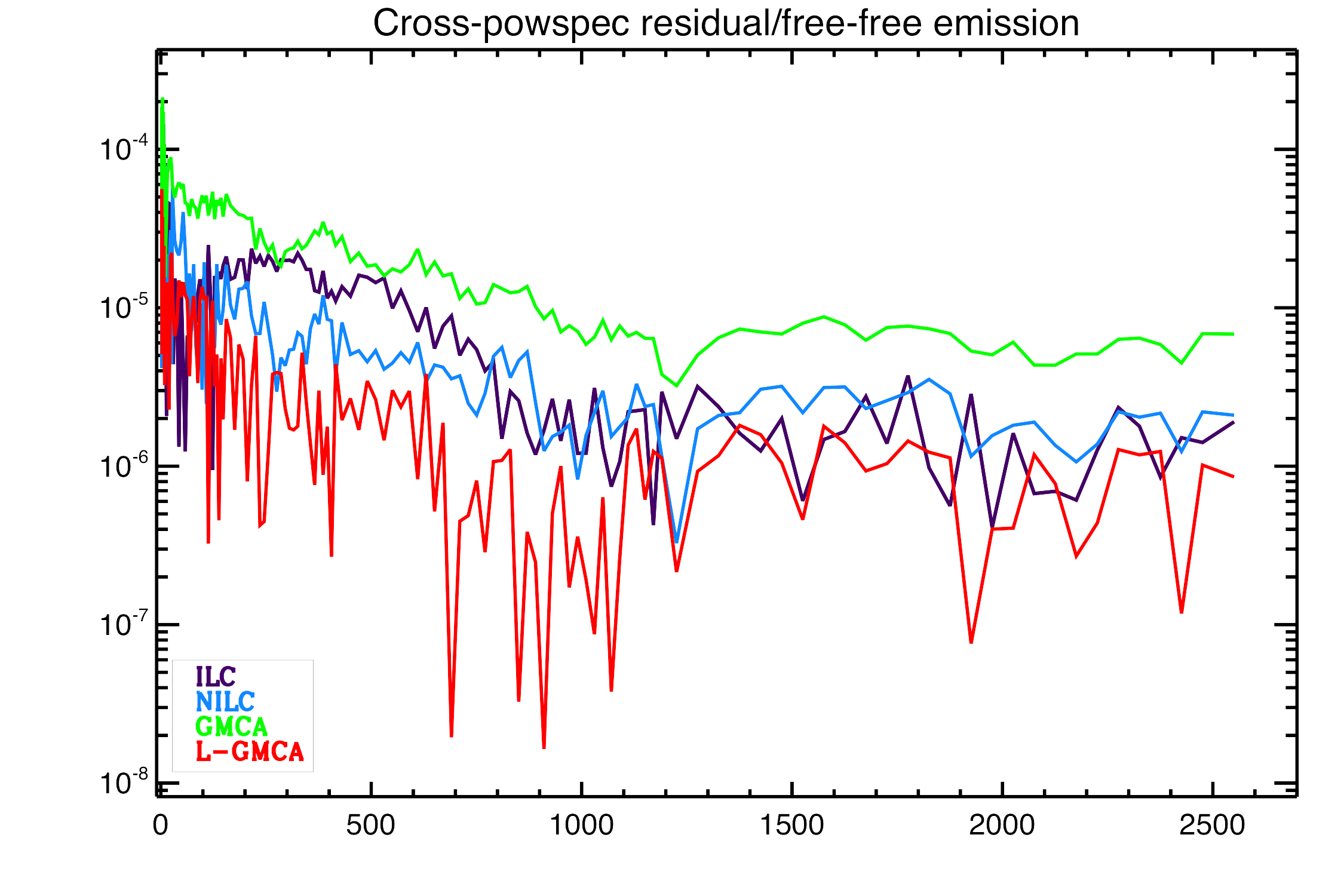} 
\end{tabular}
\caption{Cross-power spectra of the estimated CMB map with the free-free template.} 
\label{fig:ffree_ct}
\end{figure}

\begin{figure}[h!]
\begin{tabular}{c}
\includegraphics[scale=0.2]{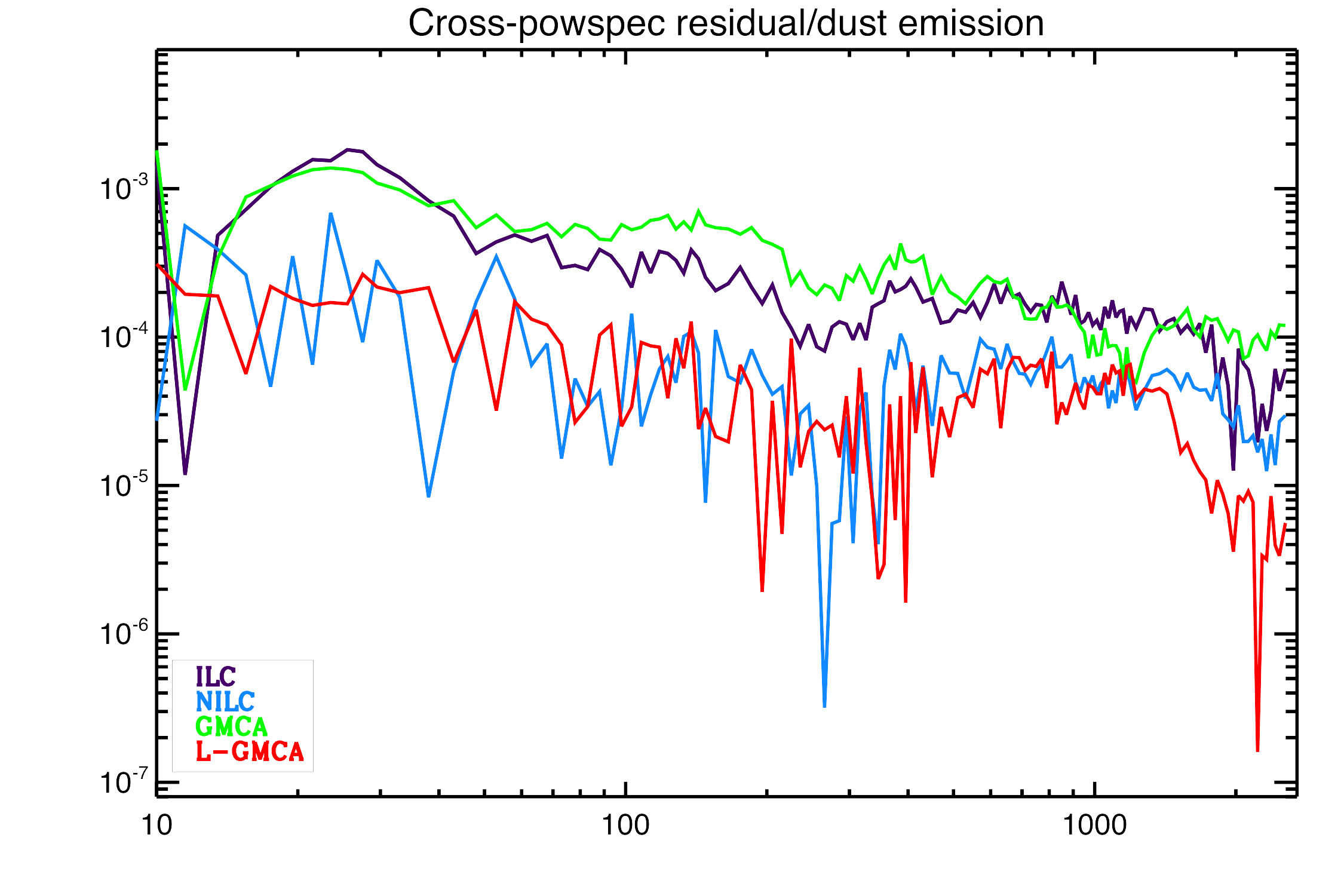} \\
 \includegraphics[scale=0.2]{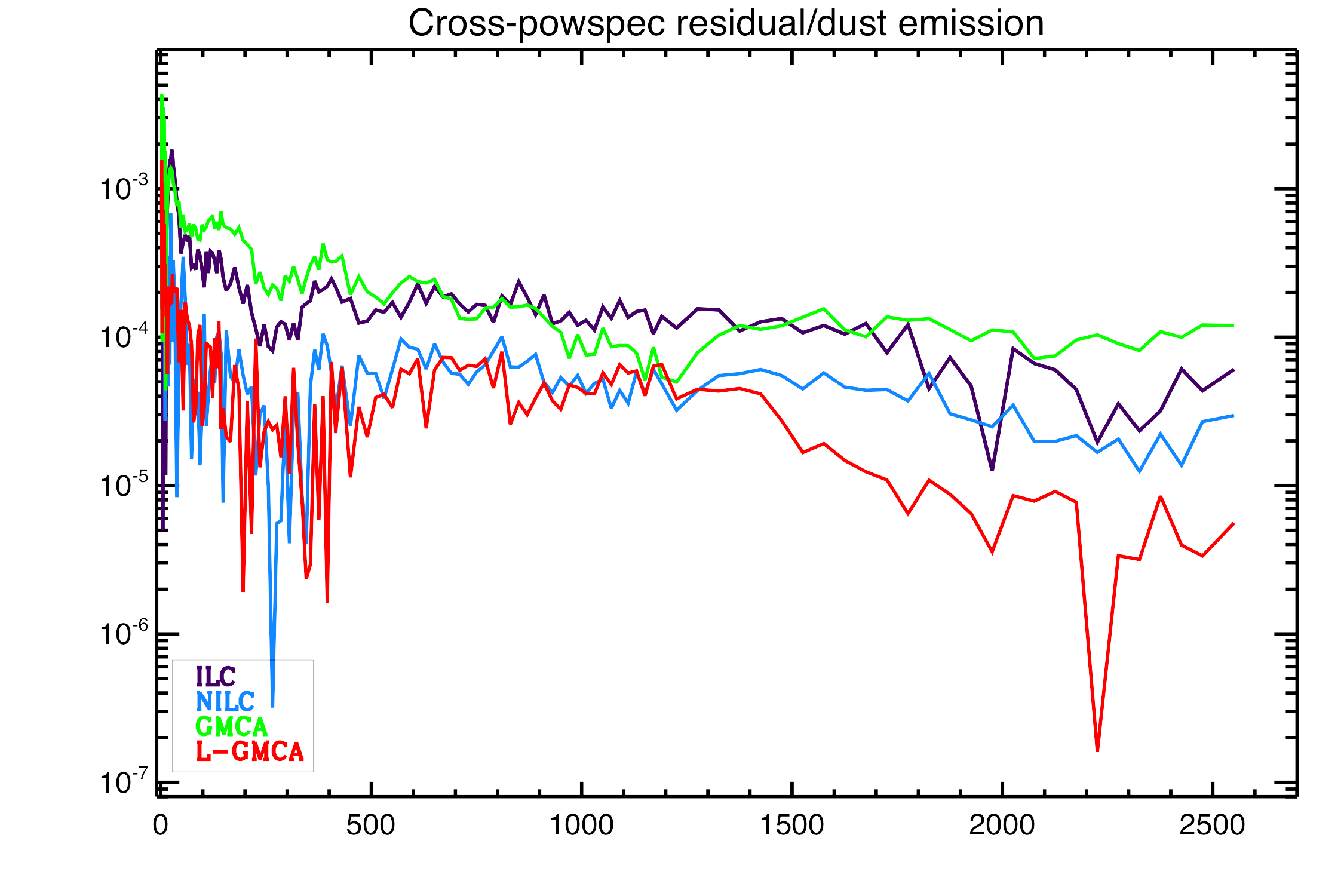} 
\end{tabular}
\caption{Cross-power spectra of the estimated CMB map with the thermal dust template.} 
\label{fig:dust_ct}
\end{figure}

\begin{figure}[h!]
\begin{tabular}{c}
\includegraphics[scale=0.2]{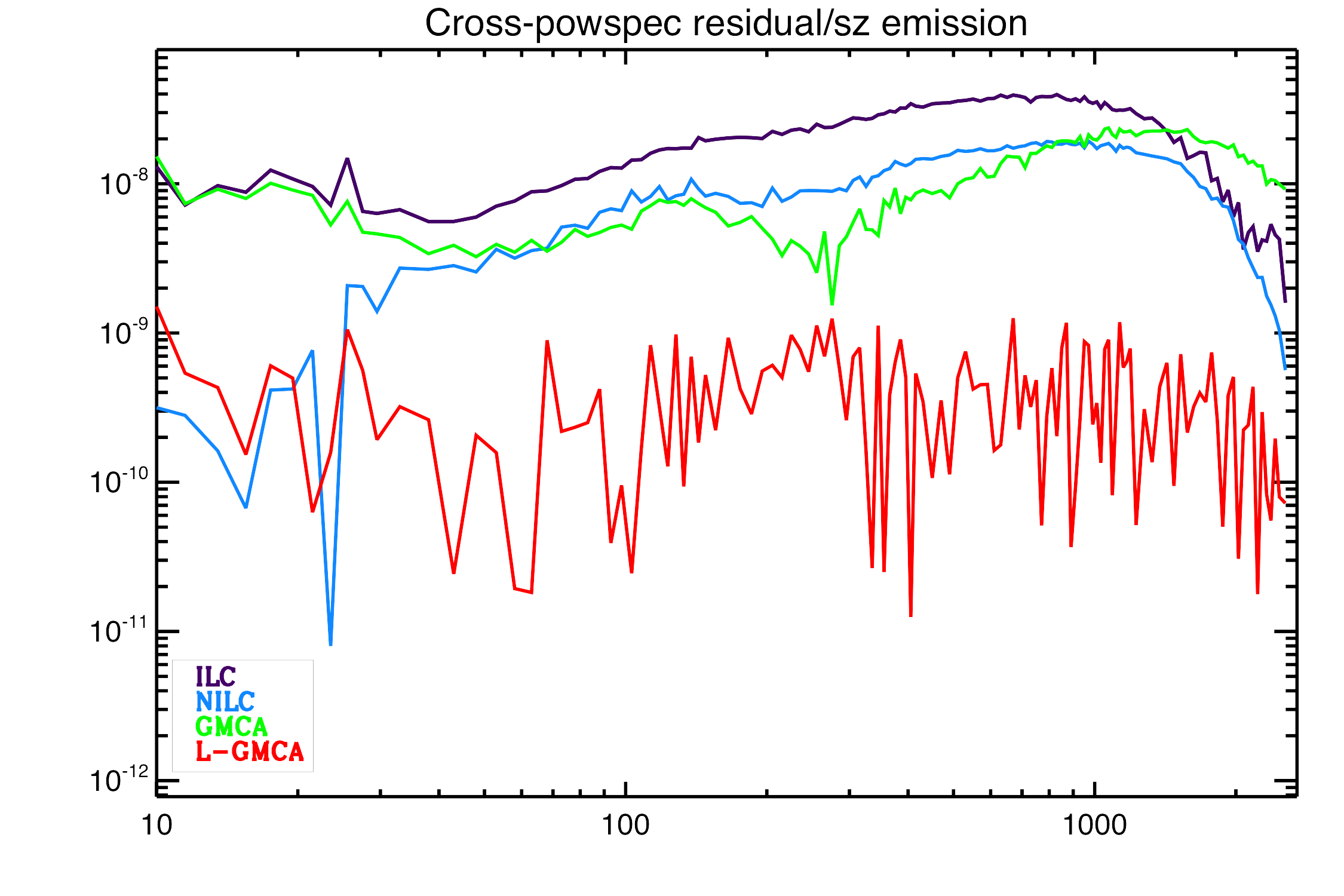} \\
 \includegraphics[scale=0.2]{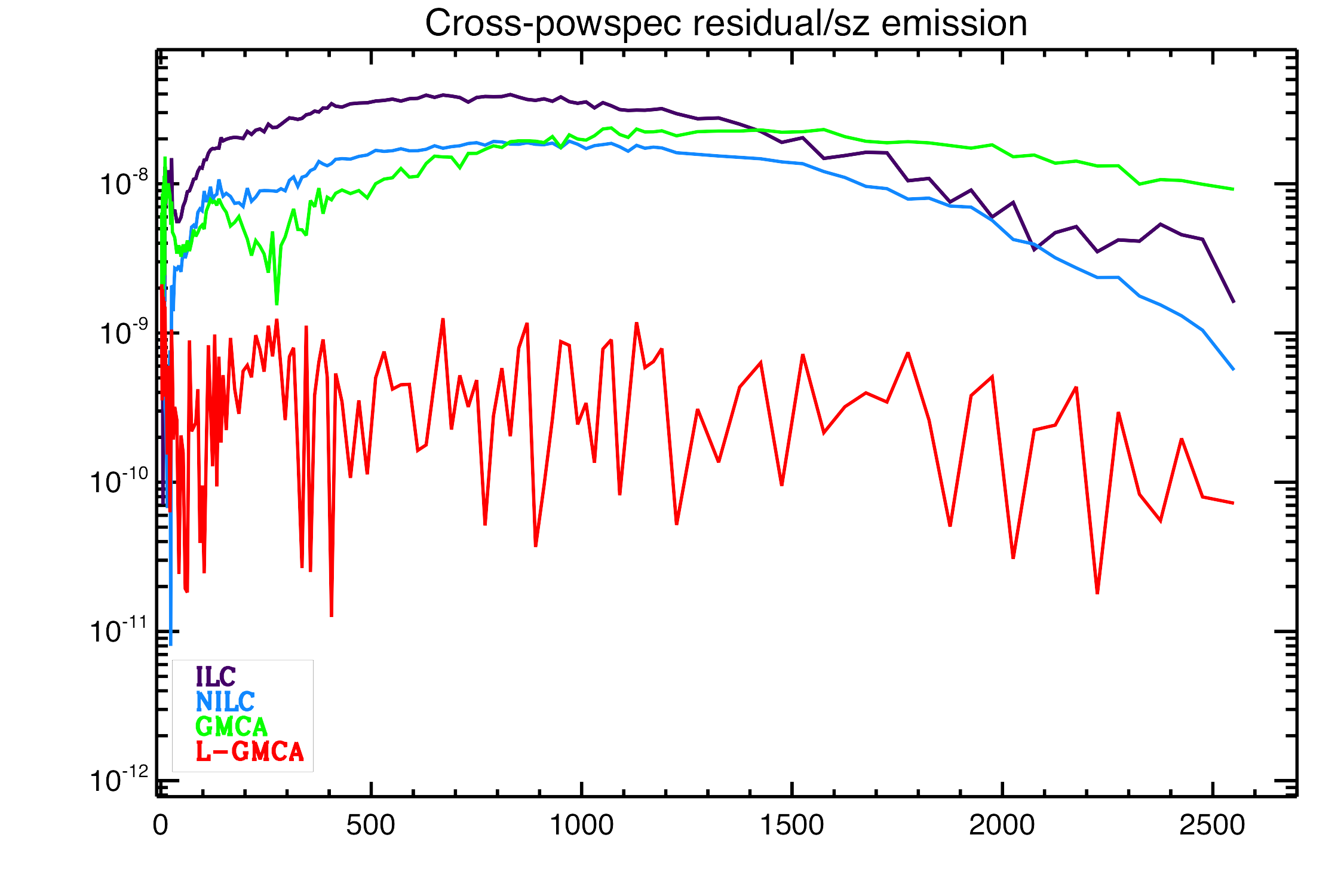} 
\end{tabular}
\caption{Cross-power spectra of the estimated CMB map with the SZ template.} 
\label{fig:sz_ct}
\end{figure}

An accurate measure of foreground contamination is the cross power spectrum between the residual map and the maps of the input foregrounds. We evaluated the cross power-spectra of foreground templates used in the simulations with the error maps defined by the difference between the input CMB map ({\it a.k.a.} the true CMB map) and estimated CMB maps. These cross power-spectra have been performed with synchrotron emission in Figure~\ref{fig:sync_ct}, Free-free in Figure~\ref{fig:ffree_ct}, dust emission in Figure~\ref{fig:dust_ct} and SZ effect in Figure~\ref{fig:sz_ct}. The same study has been carried out with synchrotron but it did not show significant differences between the four methods.\\


The conclusion we can draw from this cross-correlation study are the following~: \\
\begin{enumerate}
\item{\bf Accurately accounting for the beams~:} The heterogeneity of the data ({\it i.e.} the beam varies across frequencies) is better accounted for. As explained in Section~\ref{sec:beams}, an inaccurate modeling for the beams increases the complexity of the mixtures; even if the emission law of a given component is spatially invariant, the variability of the beam across frequencies make this emission law variant across multipoles in spherical harmonics. This is especially striking for the SZ component in Figure~\ref{fig:sz_ct}~: its emission law is exactly known and fixed in both GMCA and L-GMCA. However the two methods show radically different cross-correlation with the SZ template. Fixing the SZ emission does not guarantee that its contamination level vanishes if the beam is not properly taken into account.\\
In these simulations, the electromagnetic spectrum does not vary across the sky; assuming the beam does not change across frequencies, its estimation should be made efficiently at all scales with methods based on global mixture models. However, as shown in Figure~\ref{fig:ffree_ct}, GMCA and ILC exhibit a high level of Free-free contamination; especially for $\ell > 100$ in the case of ILC. Again, the ability of the proposed modeling to account for heterogeneous beams allows for a lower free-free contamination level.\\ \\
\item{\bf A more flexible modeling~:} whether NILC or L-GMCA, the local and multiscale mixture model we introduced in this paper allows for more degrees of freedom to better analyze complicatedly mixed components. In all the correlations computed so far, the techniques based on this modeling have outperformed the methods using the simple (but commonly used) global linear mixture model. This is particularly the case for synchrotron, Free-free and dust emissions in Figure~\ref{fig:sync_ct}, \ref{fig:ffree_ct} and Figure~\ref{fig:dust_ct} where methods based on the local/multiscale model exhibit lower foreground contamination.\\ \\
\item{\bf Sparsity versus second-order statistics~:} the large SZ contamination of ILC-based methods enlightens the low efficiency of second-order statistics to capture non-gaussian foregrounds. Conversely, sparsity-based component separation techniques are much more effective at separating the CMB and non-gaussian contaminants. The sensitivity of sparsity to higher-order statistics is very likely at the origin of the lower synchrotron and dust contamination of L-GMCA for $\ell > 1000$. \\
\end{enumerate}

%
%

\subsection{Non gaussian contamination}

\begin{figure*}[htb]
\centerline{\includegraphics[scale=0.2]{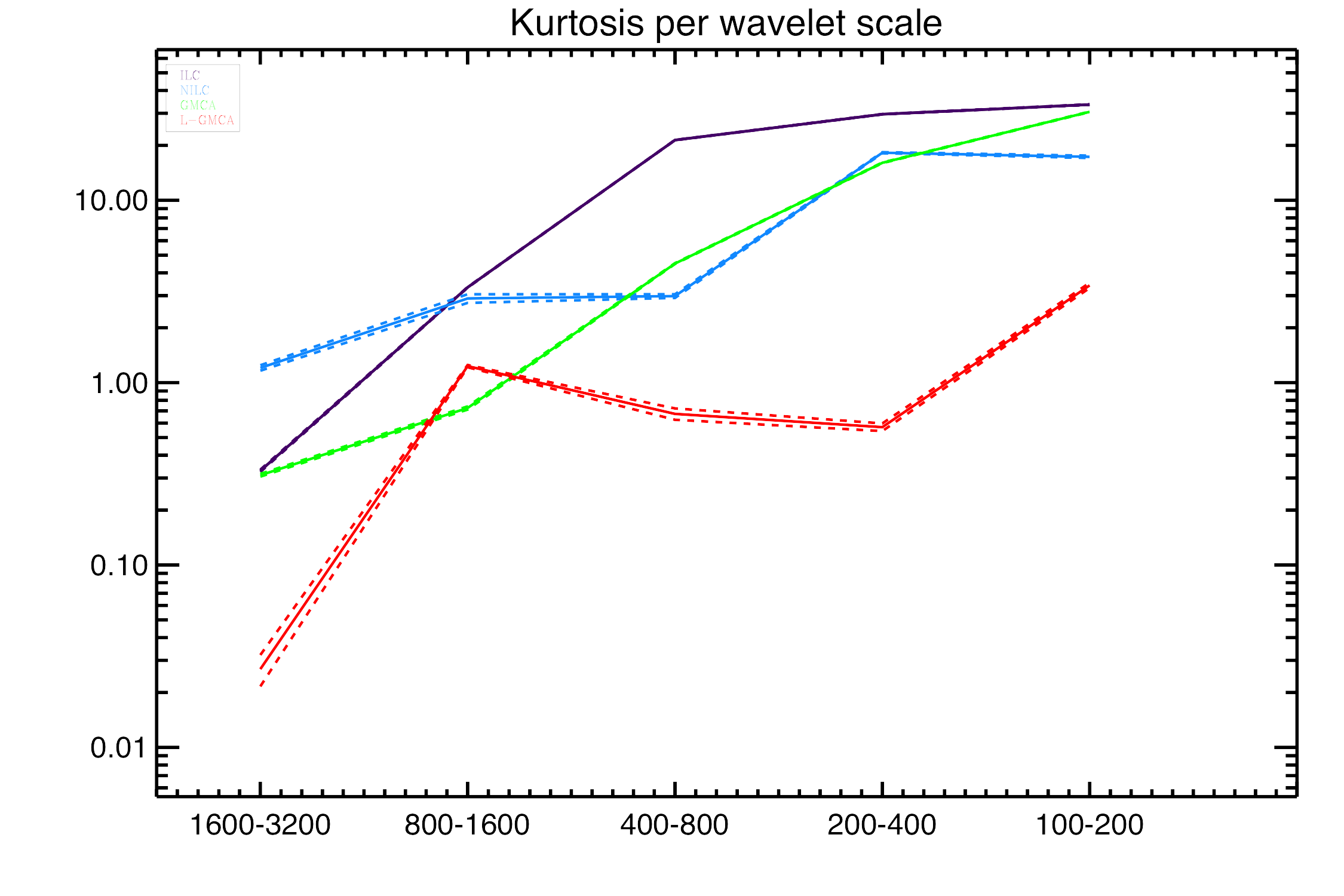}}
\begin{tabular}{cc}
\includegraphics[scale=0.2]{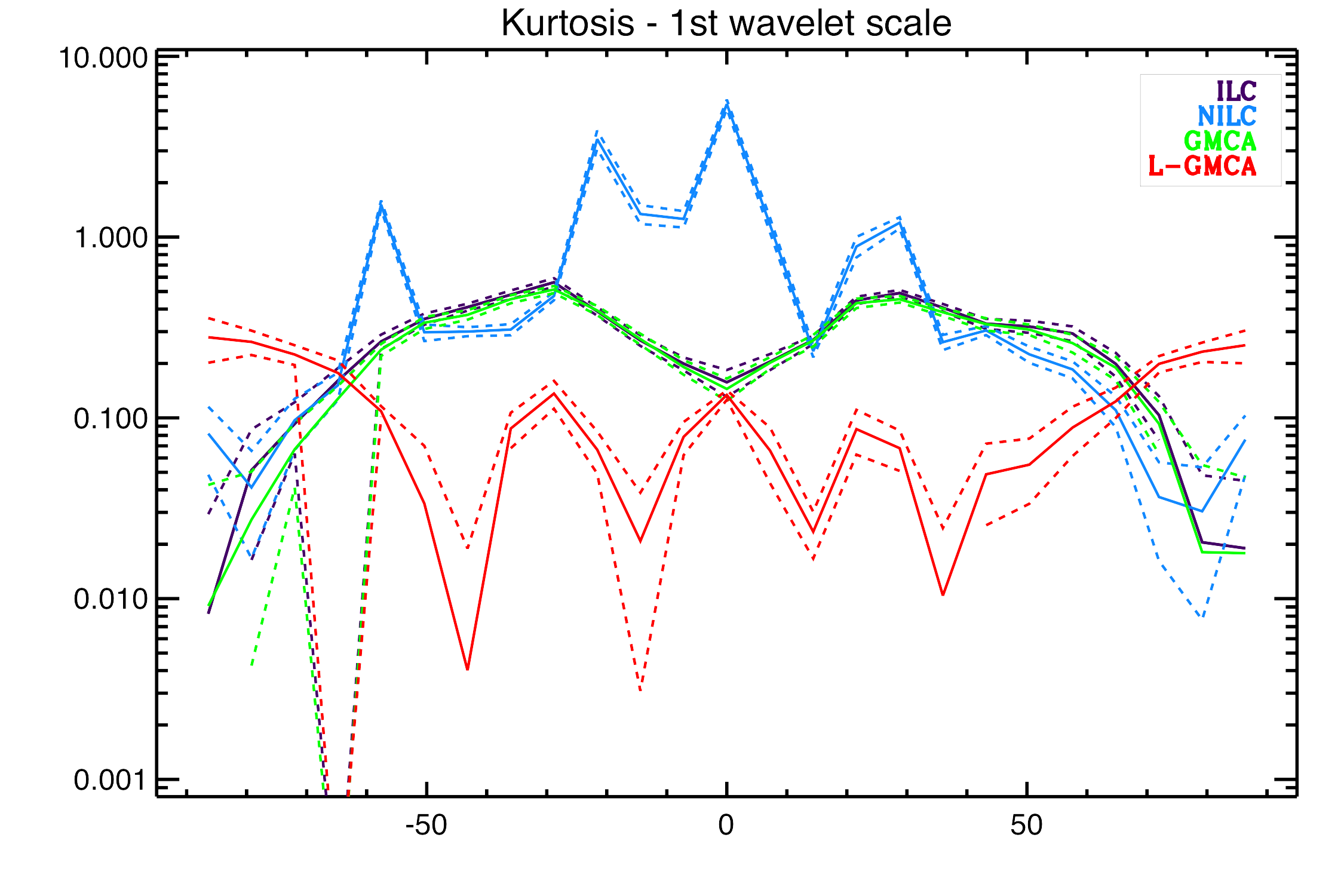} & \includegraphics[scale=0.2]{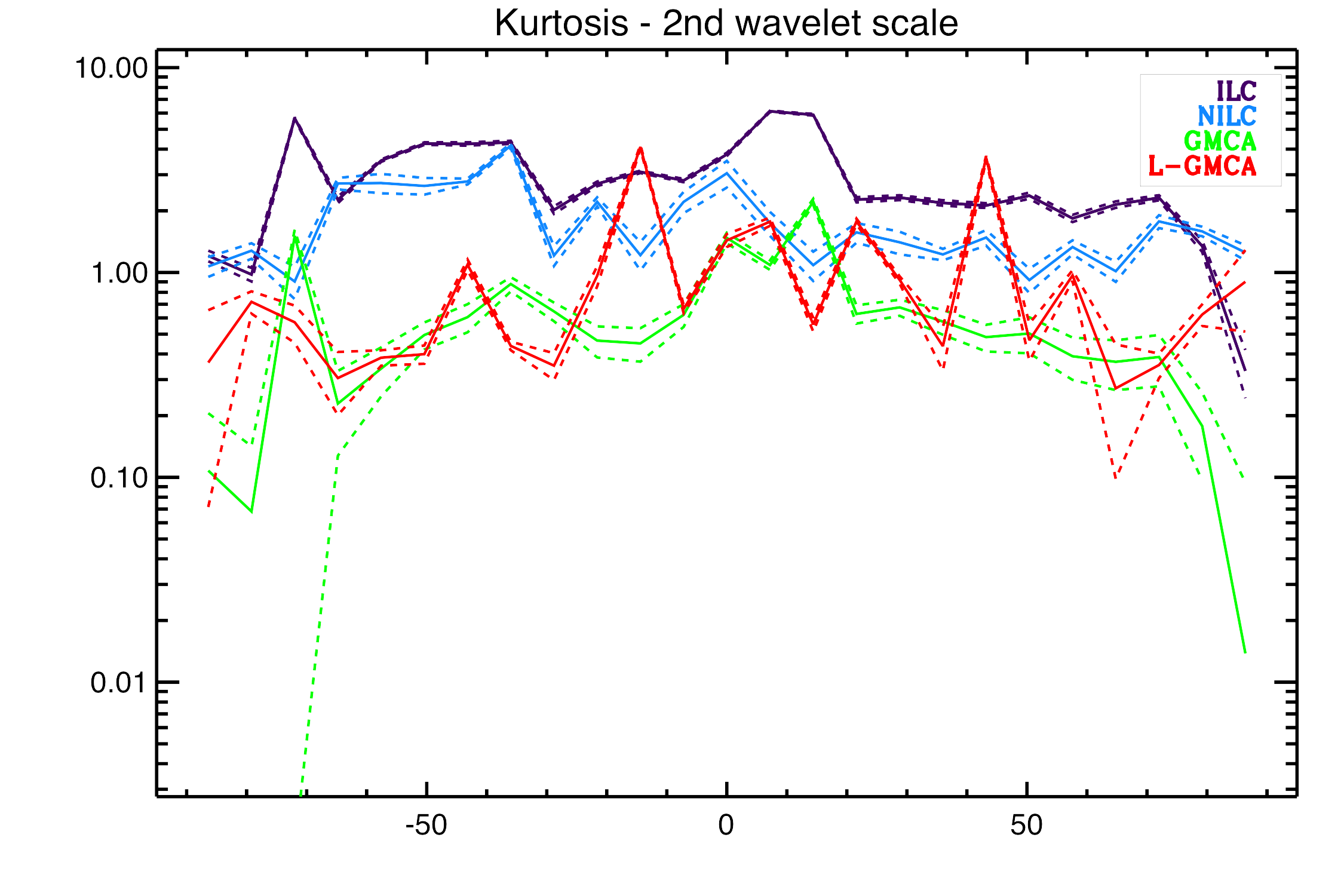} \\
\includegraphics[scale=0.2]{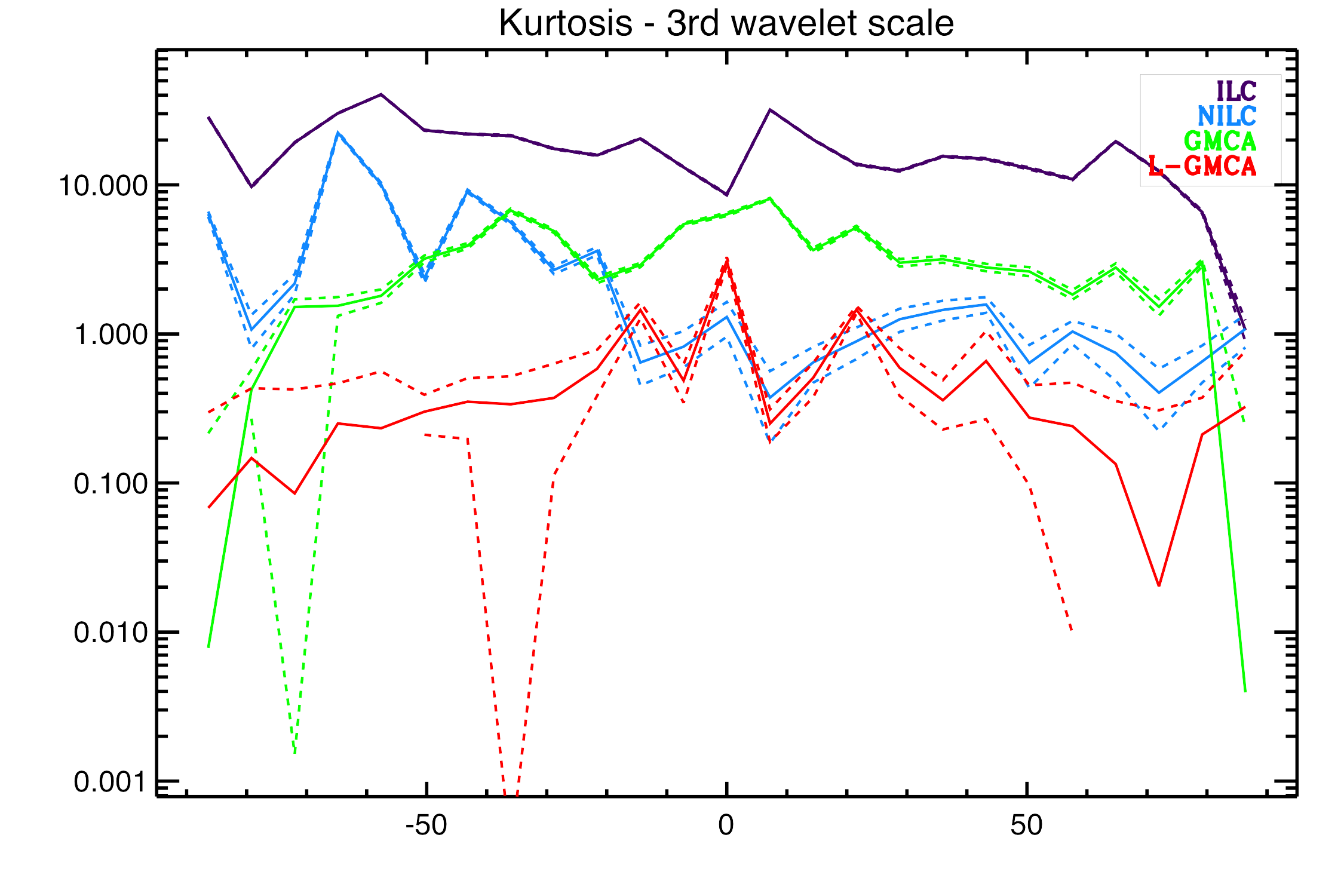} & \includegraphics[scale=0.2]{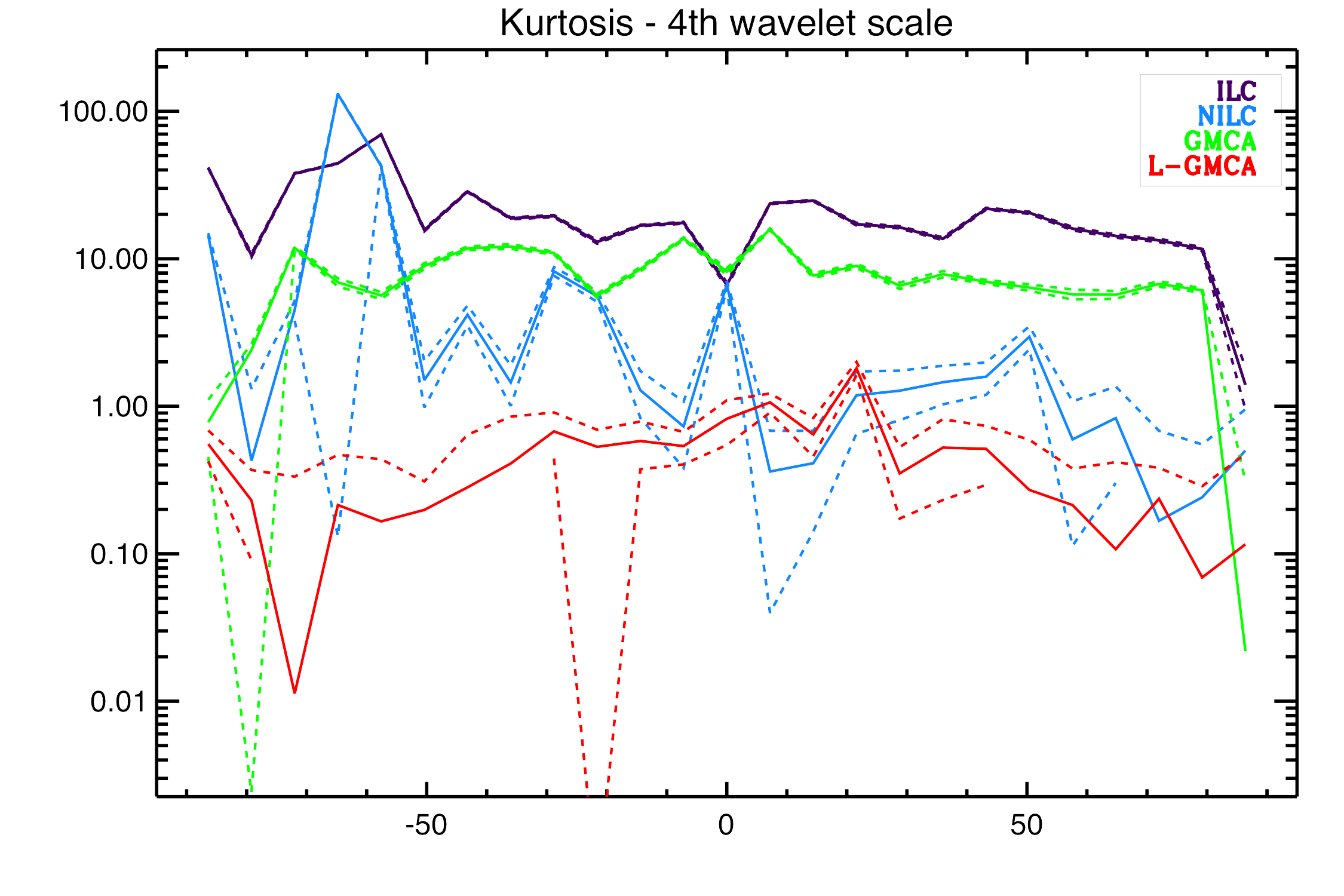} \\
\end{tabular}
\caption{Kurtosis of the estimated CMB maps in the wavelet domain. {\it Top~:} per wavelet scale. Per latitude - {\it top-left panel~:} first wavelet scale. {\it top-right panel~:} second wavelet scale. {\it bottom-left panel~:} third wavelet scale. {\it bottom-right panel~:} fourth wavelet scale. Dashed lines are $3\sigma$ confidence interval computed from $25$ random noise realizations.} 
\label{fig:kurt}
\end{figure*}

The CMB we used in these simulations is a perfect correlated Gaussian random field; precisely, it contains no trace of non-gaussianity whether it may be ISW, lensing or $\mbox{f}_{NL}$. CMB non-gaussianities will evidently originate form spurious foreground contaminations. In this paragraph, we propose evaluating the level of (non) Gaussianity of the estimated CMB. The CMB being perfectly Gaussian in these simulations, such a study will give a different measure of the contamination level.\\
Common non-gaussianity tests consist in computing the higher-order statistics of the estimated residual maps in the wavelet domain \citep{starck:book06}. The aim of this study is to give a different way to measure foreground contamination. Therefore, we rather opted for the evaluation of the  non-gaussianity (NG) level of the residual maps instead of the CMB. This has the crucial advantage of providing a contamination level measure insensitive to CMB fluctuations.\\
For this purpose, several statistical tests were performed~: the kurtosis ({\it i.e.} statistics of order $4$) is calculated in the $5$ first wavelet scales of the residual maps. These results are shown in the top pictures of Figure~\ref{fig:kurt}. Except at very large scale ($\ell < 200$), these two measures of non-gaussianities draw very similar conclusion~: i) in the first wavelet scale, all the three methods are likely to be compatible with a Gaussian contamination; this largely originates from the preeminence of noise in this range. ii) the sparsity -based separation criterion used in L-GMCA yields lower NG levels in the scales $2$ to $4$ ({\it i.e.} $200 \leq \ell \leq 1600$). At large scale, L-GMCA has a lower kurtosis value. At that stage, massive simulations of all the components (and not only noise) would be mandatory to precisely evaluate the performances of these methods at low $\ell$.\\
The bottom graphics of Figure~\ref{fig:kurt} features the value of the kurtosis in fixed wavelet scales but in different equi-angular bands of latitude. By convention, the data were in galactic coordinates; latitude $0$ thus corresponds to the galactic plane. From these pictures, L-GMCA is likely to be less contaminated than the other three residual maps. More significantly, GMCA exhibits lower NG contamination at all latitudes in the scales $2$ to $4$. This enlightens the role of GMCA and more precisely the use of sparsity to better extract non-gaussian sources. It is very likely that the second order statistics used as a separation criterion in ILC is less sensitive that sparsity to separate foregrounds. This unveils the positive impact of the local and multiscale model together with a sparsity-based separation criterion as used in L-GMCA.

\section{Discussion - incorporating astrophysical models}
\label{sec:discuss}

Except for CMB and SZ, no precise astrophysical model is currently accounted for within GMCA. Remarkably, incorporating more precise modeling of some of the components would mainly consist in adding up additive constraint to the problem in Equation~\ref{eq:GMCA2}. In \citet{Leach_08}, a model-based extension of GMCA was also partly evaluated. More precisely the "model GMCA" modeled the emission law of the synchrotron and free-free emission as power laws with fixed but unknown spectral index. Interestingly, the CMB map estimate we obtained at that time seemed particularly less contaminated by these low frequency foregrounds at the expense of a significantly large noise level. However, as shown in this paper, the first attempt to model for more sophisticated astrophysical {\it prior} models were kind of useless efforts as long as the variation of the beam across scale was not properly accounted for. The model introduced in this paper should allow for a more effective implementation of these models~: i) the beam issue is now properly tackled, ii) the local/multiscale model offers a privileged framework to model for the natural multiscale variation of models' parameters such as the spectral index of the synchrotron emission and the temperature/spectral index of the thermal dust emission. Beyond models, the use of ancillary data ({\it e.g.} Haslam map for the synchrotron emission, IRAS map for the thermal dust emisson, H$\alpha$ map for the free-free emission) within GMCA should be possible. It is not clear yet how to proceed with a extension of L-GMCA to a model-based component separation techniques. Two approaches could be envisioned~: i) implementing astrophysical modeling to jointly estimate the components' maps as well as model parameters at the expense of an increase of the numerical complexity of the estimation problem, ii) alternating between L-GMCA with fixed parameters for the foregrounds and a template fitting/parameter estimation for parameter estimation. This is out of the scope of this study and is reserved for future work.

\section*{Software}
Source separation techniques used in cosmology and more precisely for CMB map estimation are generally high-end method which are seldom publicly shared and available online. 
The L-GMCA code will be made publicly available at {\it http://www.cosmostat.org/lgmca.html}.

\section{Conclusion}
\label{sec:conc}
The estimation of a high precision CMB map featuring low noise and low foreground contamination is of crucial interest for the astrophysical community. This problem is customarily tackled in the framework of component separation. As any estimation problem, the accurate modeling of the data is essential. However, a close look at the astrophysical phenonema at play in the CMB data such as WMAP or Planck reveals that the linear mixture model used so far by common component separation methods in cosmology does not hold. First, the variation of the beam across scale is seldom accounted for which highly limits the performances of these component separation methods especially at small scales. More importantly, the linear mixture model does not afford enough degrees of freedom to precisely capture the complexity of the data including the variability across space of the emission law of some of the components of interest. To alleviate these limitations, this paper introduces a new modeling of the components' mixtures using a multiscale and local decomposition of the data in the wavelet domain. We introduced a novel sparsity-based coined L-GMCA (Local - Generalized Morphological Component Analysis) which makes profit of the proposed local/multiscale mixture model. In the proposed framework, wavelet-based multiscale analysis allows for a precise modeling of the beam evolution across channels. Capturing the variations across pixels of the emissivity of components is carried out by partitioning each wavelet scale with adaptive patch sizes. Extensive numerical experiments have been carried out which show the superiority of the proposed modeling and separation technique to provide a clean, low-foreground CMB map estimate. More precisely, we showed that the local and multiscale modeling allows for improved separation results whether it is used with separation techniques as different as ILC and GMCA. Additionally, the numerical experiments enlightens the dramatic positive impact of the use of sparsity in L-GMCA to provide less galactic foreground contamination as well as significantly lower non-gaussianity levels.

\section*{Acknowledgment}
This work was supported by the French National Agency for Research (ANR -08-EMER-009-01) and 
the European Research Council grant SparseAstro (ERC-228261) .

\appendix

\section{Multiscale local mixture model}
\label{sec:lmm2}
Before going further into the description of the method, principles of the multichannel quadtree decomposition and notations have to be discussed. In multiscale image analysis, quadtree decomposition amounts to decomposing each wavelet scale in patches with dyadic sizes starting from the original field itself and sequentially subdivide each patch in $4$. To our knowledge, this multiscale analysis procedure has never been extended to analyze multichannel data in a source separation framework. In this specific context, such an extension consists in estimating the mixture parameters (\textit{i.e.} the mixing matrix $\bf A$ and the sources $\bf S$) at each scale $\mu$ on patches with various patch sizes starting from large patches of size $2^{L_\mu} p_\mu$ to smaller patches of size $p_\mu$. The parameter $L_\mu$ denotes the number of decomposition levels or subdivisions in which the estimation of the mixture parameters is carried out.\\
Figure~\ref{fig:quadt} clearly illustrates the principle of the decomposition~: the analysis ({\it i.e.} estimation of the mixing matrices) is first performed on the largest patch size allowed $2^{L_\mu} p_\mu$~: ${\bf W}^{(\mu)}_{L_\mu}[k]$. In the next step, the patch ${\bf W}^{(\mu)}_{L_\mu}[k]$ is divided into $4$ non-overlapping patches of size $2^{L_\mu-1} p_\mu$ on which the parameters estimation is carried out. The same process is performed until the final patch size $p_\mu$ is attained.\\
The notation we will be using in the sequel is the following~: 
\begin{itemize}
\item{\bf Multichannel patch ${\bf W}^{(\mu)}_l[k]$~:} Is the concatenation of $N_\mu$ frequency patches of size $2^l p_\mu \times 2^l p_\mu $ as defined in Figure~\ref{fig:quadt} centered on pixel $k$. We invite the reader to notice that the three indices of the notations ${\bf W}_l^{(\mu)}[k]$ underlines the dependence of this variable on the~: i) the pixel at which the analysis is carried out, ii) the wavelet scale $\mu$ and quadtree decomposition level $l$.
\item{\bf Multichannel patch ${\bf A}^{(\mu)}_l[k]$~:} Mixing matrix estimated from the multichannel patch ${\bf W}_l^{(\mu)}[k]$.
\end{itemize}

Interestingly, the same area of the sky being analyzed at different mixture scales (\textit{i.e.} patch sizes), it makes it possible to choose afterwards the "optimal'' patch size from the different estimates obtained for $l = 0,\cdots,L_\mu$. It therefore allows for more degrees of freedom to select an adapted patch size at each location.\\

Full-sky CMB data ({e.g.} WMAP, Planck) are generally sampled on the Healpix sampling grid \citep{Healpix}. For the sake of simplicity and computational efficiency, the most straightforward way to decompose an Healpix map into patches is to decompose each Healpix face.


\subsection*{Choosing the "optimal" patch decomposition}
\label{sec:localestim}

For each patch location $k$, L-GMCA provides a sequence of parameters or more specifically mixing matrices computed at different mixtures scales~: $\left\{ {\bf A}^{(\mu)}_l[k] \right\}_{l=0,\cdots,L_\mu}$. These mixing matrices have been computed from patches of increasing sizes; this entails that they offer different views of the data in the neighborhood of pixel $k$. As emphasized in the previous section, this allows for extra flexibility to choose adapted local parameters~: i) mixing matrices estimated from larger areas will be more adapted to stationary areas, ii) mixing matrices evaluated from local patches analysis are well-suited to better capture local variations of the foreground emissivity. The difficulty now mainly consists in defining a strategy to select the {\it optimal} local estimator of the mixing matrix (\textit{i.e.} optimal amongst the set of available estimators).\\
First, it has to be noticed that in the framework of GMCA and similarly in L-GMCA, once the mixing matrices is estimated, the final estimate of the components are computed by applying the Moore pseudo-inverse of the mixing matrix to the data~:
$$
{\bf S}^{(\mu)} = {{\bf A}^{(\mu)}}^+ {\bf W}^{(\mu)}
$$
This way the sources {\bf linearly} depend on the original data ${\bf W}^{{\mu}}$; this is particularly important to control error propagation in the final CMB map estimate.\\
Discriminating between the available mixing matrices at location $k$ amounts to defining a criterion sensitive the a {\it local} mis-estimation of the CMB map. From the sequences of available mixing matrices $\{{\bf A}^{(\mu)}_l[k]\}_{l=0,\cdots,L_\mu}$, one can compute a sequence of various local CMB map estimates about pixel $k$ at the level of the smallest patch of size $p_\mu$. These estimates are computed by applying the pseudo-inverse of the different matrices $\{{\bf A}^{(\mu)}_l[k]\}_{l=0,\cdots,L_\mu}$ to the smallest data patch about pixel $k$, ${\bf W}^{(\mu)}_{l=0}[k]$~:
 $$
 {\bf S}^{(\mu)}_l[k] =  {{\bf A}_l^{(\mu)}}^+ {\bf W}_{l=0}^{(\mu)}[k] 
 $$ 
Quite naturally, the optimal mixing matrix amongst the available estimators $\left\{ {\bf A}_l^{(\mu)}[k] \right\}_{l=0,\cdots,L_\mu}$ should be chosen such that the estimated CMB  is the least contaminated possible. Here contamination is taken in the wide sense: foreground and noise. One remarkable property of the CMB is its decorrelation with the instrumental noise and all other astrophysical components. More quantitatively, owing the linearity of the estimation process, the estimated CMB can be modeled as~:
$$
s^{(\mu)} = {s^{(\mu)}}^\star + r^{(\mu)} + n^{(\mu)}
$$
where $s^{(\mu)}$ stands for the estimated CMB in scale $\mu$, ${s^{(\mu)}}^\star$ the sought after clean CMB, $r^{(\mu)}$ the residual coming from the foregrounds contaminants and $n^{(\mu)}$ the instrumental noise. As the CMB is decorrelated with both the residual and noise, the variances of the three terms in the above equation add up~:
$$
\mbox{Var} \left \{s^{(\mu)} \right \} = \mbox{Var} \left \{{s^{(\mu)}}^\star \right \} + \mbox{Var} \left \{r^{(\mu)}\right\} + \mbox{Var} \left \{ n^{(\mu)} \right\}  
$$
As a consequence, a solution with higher noise or foreground contamination will have a higher variance. It is then natural to choose the mixing matrix amongst the sequence $\left\{ {\bf A}^{(\mu)}_l[k] \right\}_{l=0,\cdots,L_\mu}$ which yields the local CMB estimate with the lowest variance. This suggests that the {\it optimal} mixing matrix and local CMB map should be chosen as follows~:
\begin{equation}
\begin{split}
l^{(\mu)}_\star[k] = \mbox{Argmin}_{l = 0,\cdots,L_\mu} \, \mbox{Var}\left\{ s^{(\mu)}_{l=0}[k] \right\} \\
\mbox{ where } s^{(\mu)}_{l=0}[k] = \left[{{\bf A}^{(\mu)}_l[k]}^+ {\bf W}^{(\mu)}_{l=0}[k]\right]_{1}
\end{split}
\end{equation}
where the operator $\left[ \, . \, \right]_{1}$ extracts the CMB out of the entire set of estimated components.\\
Other estimator selection criteria can be envisioned based in statistical characteristics of the CMB map such as its gaussianity. Foreground contamination are very likely to increase the non-gaussianity level of the estimated CMB map. Non-gaussianity measures such as higher-order cumulants could be used to discriminate between various local estimates of the CMB. However, these measures would be less sensitive to the noise contamination.

\bibliographystyle{aa}
\bibliography{biblio}

\end{document}